\documentclass{aa} 
\usepackage{graphicx}
\usepackage{txfonts}
\usepackage{multirow}
\usepackage[shortlabels]{enumitem}
\usepackage{hyperref}
\graphicspath{{./}{Figures/}}

\begin{document}

   \title{Detectability of biosignatures on LHS~1140~b}
             
    \author{Fabian Wunderlich\inst{1,2}
    \and Markus Scheucher\inst{1,2}
    \and John Lee Grenfell\inst{2}
   \and Franz Schreier \inst{3}
   \and Clara Sousa-Silva \inst{4}
   \and Mareike Godolt\inst{2}
   \and Heike Rauer\inst{1,2,5}}

\institute{Zentrum für Astronomie und Astrophysik, Technische Universität Berlin, Hardenbergstraße 36, 10623 Berlin, Germany\\ \email{fabian.wunderlich@dlr.de}
  \and Institut für Planetenforschung, Deutsches Zentrum für Luft- und Raumfahrt, Rutherfordstraße 2, 12489 Berlin, Germany 
  \and Institut für Methodik der Fernerkundung, Deutsches Zentrum für Luft- und Raumfahrt,  82234 Oberpfaffenhofen, Germany
  \and Center for Astrophysics, Harvard \& Smithsonian, 60 Garden Street, Cambridge, MA 02138, US
  \and Institut für Geologische Wissenschaften, Freie Universität Berlin, Malteserstr. 74-100, 12249 Berlin, Germany}

   \date{}

% \abstract{}{}{}{}{} 
% 5 {} token are mandatory
 
  \abstract
   {Terrestrial extrasolar planets around low-mass stars are prime targets when searching for atmospheric biosignatures with current and near-future telescopes. The habitable-zone Super-Earth LHS~1140~b could hold a hydrogen-dominated atmosphere and is an excellent candidate for detecting atmospheric features.}
   {In this study, we investigate how the instellation and planetary parameters  influence the atmospheric climate, chemistry, and spectral appearance of LHS~1140~b. We study the detectability of selected molecules, in particular potential biosignatures, with the upcoming James Webb Space Telescope (JWST) and Extremely Large Telescope (ELT).}
   {In a first step we use the coupled climate-chemistry model, 1D-TERRA, to simulate a range of assumed atmospheric chemical compositions dominated by molecular hydrogen (H$_2$) and carbon dioxide (CO$_2$). 
   Further, we vary the concentrations of methane (CH$_4$) by several orders of magnitude.
   In a second step we calculate transmission spectra of the simulated atmospheres and compare them to recent transit observations. Finally, we determine the observation time required to detect spectral bands with low resolution spectroscopy using JWST and the cross-correlation technique using ELT.}
   {In H$_2$-dominated and CH$_4$-rich atmospheres oxygen (O$_2$) has strong chemical sinks, leading to low concentrations of O$_2$ and ozone (O$_3$). The potential biosignatures ammonia (NH$_3$), phosphine (PH$_3$), chloromethane (CH$_3$Cl) and nitrous oxide (N$_2$O) are less sensitive to the concentration of H$_2$, CO$_2$ and CH$_4$ in the atmosphere. In the simulated H$_2$-dominated atmosphere the detection of these gases might be feasible within 20 to 100 observation hours with ELT or JWST, when assuming weak extinction by hazes. } 
   {If further observations of LHS~1140~b suggest a thin, clear, hydrogen-dominated atmosphere, the planet would be one of the best known targets to detect biosignature gases in the atmosphere of a habitable-zone rocky exoplanet with upcoming telescopes. }

   %\keywords{giant planet formation --
%                $\kappa$-mechanism --
%                stability of gas spheres
%               }

   \maketitle
%
%-------------------------------------------------------------------

\section{Introduction}
The nearby temperate Super-Earth LHS~1140~b \citep{dittmann2017,ment2019} is an exciting target for atmospheric characterization. \citet{morley2017} assumed Venus, Titan and Earth-like atmospheres for LHS~1140~b. Their results suggest that atmospheric characterization with the James Webb Space Telescope (JWST) could be possible although challenging. 

A recent study of the temporal radiation environment of the LHS 1140 system
suggests that the planet receives relatively constant near ultraviolet (NUV)
(177-283~nm) flux <2\% compared to that of Earth \citep{spinelli2019}. 
The results of \citet{chen2019} suggest that LHS 1140~b might be stable
against complete ocean desiccation due to the low UV activity
of the host star, which would bode well for its habitability. However,  \citet{chen2019} assumed a rather low UV flux for the star which might lead to an underestimation of the water loss. Further, due to the extended pre-main sequence phase of M dwarfs \citep[see e.g.][]{baraffe2015,luger2015} LHS 1140~b may have experienced extreme water loss before the star entered the main sequence phase \citep[see e.g.][]{luger2015}.

Assuming an Earth-like atmosphere with updated sea-ice paramerization the 3D model study of \citet{yang2020} suggested a reduced surface ocean on LHS~1140~b (from 12\% to 3\% surface coverage). \citet{diamond2020} observed two transits of LHS~1140~b with the twin Magellan Telescopes but their analysis suggested that a precision increased by a factor of about 4 was needed  for the detection of e.g. a cloudless hydrogen atmosphere present at amounts consistent with the bulk density. Recently, \citet{edwards2020} presented spectrally resolved observations of LHS~1140~b using the G141 grism of the Wide Field Camera 3 (WFC3) on the Hubble Space Telescope (HST). Their results suggest that the planet may host a clear H$_2$-dominated atmosphere and show evidence of an absorption feature at 1.4~$\mu$m. 
%However, such an feature also might be produced by CH$_4$ absorption \citep[see][]{bezard2020,blain2020}

The processes affecting climate and composition of Super-Earths, such as LHS 1140~b are not well known. 
Evidence was found that there is a dip in the radius distribution of extrasolar planets at 1.5--2.0 $R_{\oplus}$ \citep[see e.g.][]{owen2013, fulton2017, van-eylen2018, hardegree2020}. With a radius of $\sim$1.7 $R_{\oplus}$ LHS~1140~b lies within this so called 'Radius Valley' \citep{ment2019}, which is interpreted as the transition between predominantly rocky planets and volatile-rich planets. A number of studies have investigated the origin of the Radius Valley \citep[see e.g.][]{owen2013, lee2014, owen2017,lopez2018,ginzburg2018,gupta2019}. 
LHS~1140~b is not expected to have a large H$_2$/He envelope due to its high bulk density of $\rho$, of 7.5$\pm$1.0 cm$^{-3}$ \citep{ment2019}. However, massive Super-Earths might retain small residual H$_2$-atmospheres at the end of the core-powered mass loss \citep[see e.g.][]{ginzburg2016,gupta2019}.

In H$_2$-dominated atmospheres significant heating could be induced by self and foreign H$_2$ Collision Induced Absorption (CIA) \citep[e.g.][]{pierrehumbert2011,ramirez2017}. Regarding composition, lessons from the solar system gas giants \citep[][and references therein]{yung1999} suggest ammonia (NH$_3$) and phosphine (PH$_3$) chemistry as well as (1) pathways starting with methane (CH$_4$) forming long chain hydrocarbons which can condense to form hazes, and (2) pathways destroying long chain hydrocarbons driven mainly by initial reaction with atomic hydrogen (H) from extreme UV (EUV) photolysis of H$_2$.
In exoplanetary science e.g. \citet{hu2014} studied photochemical responses of hydrogen atmospheres on Super-Earths; \citet{line2011} discussed processes controlling the partitioning between CH$_4$ and CO on GJ~436~b and recently \citet{lavvas2019} studied effects of photochemistry, mixing and hazes on GJ~1214~b. %(modern Earth? T_eq?) 

Clouds and hazes can obscure the observed spectrum of the planetary atmosphere below the top of the haze or cloud layer.
\citet{arney2016} and  \citet{arney2017} used a 1D climate-chemistry model to simulate the photochemically driven formation of organic hazes in the atmosphere of early Earth and exoplanets located in the habitable zones (HZs) of their host stars. 
They concluded that the concentration of CH$_4$ has a large impact on the haze formation and propose that hydrocarbon haze may be interpreted as a biosignature on planets with substantial levels of CO$_2$.

The detection of potential biosignature gases like oxygen (O$_2$), nitrous oxide (N$_2$O) or chloromethane (CH$_3$Cl) in an Earth-like or CO$_2$-dominated atmosphere will be challenging using transmission or emission spectroscopy \citep[see e.g.][]{schwieterman2018,batalha2018,wunderlich2019,lustig-yaeger2019}.
The characterization of an H$_2$-atmosphere is more favorable due to the lower mean molecular weight leading to larger spectral features. 
In such an atmosphere several potential biosignatures might be detectable including NH$_3$, dimethyl sulfide (DMS), CH$_3$Cl, PH$_3$ and N$_2$O \citep{seager2013, seager2013b, schwieterman2018, sousa-silva2020}.

In this work we apply the steady-state, cloud-free, radiative-convective photochemistry model 1D-TERRA \citep{scheucher2020,wunderlich2020} together with the theoretical spectral model GARLIC \citep{schreier2014} to simulate a range of CO$_2$, H$_2$-He atmospheres (and mixtures thereof) as well as atmospheric spectra for LHS~1140~b. 
%To our knowledge ours is the first study to perform consistent coupled climate-chemistry calculations for atmospheres dominated by H$_2$ and CO$_2$ on this potentially habitable Super-Earth. 
A central aim of our work is to investigate potential atmospheres of this Super-Earth and determine the detectability of key atmospheric features, in particular potential biosignatures, in the context of the forthcoming JWST and Extremely Large Telescope (ELT).

Section~\ref{sec:method} introduces the climate-photochemistry model 1D-TERRA, the line-by-line spectral model GARLIC, and the signal to noise (S/N) models for JWST and ELT.
In Sect.~\ref{sec:results} we first show the results of the atmospheric modelling and the resulting transmission spectra, followed by the results of the S/N calculations. We summarize and conclude our results in Sect.~\ref{sec:summary}.

%- LHS 1140b
%- H2-atmospheres
%- Detectability
%- hazes

% H2CO, HO2, CO2

% CH3 C2H6, C2H2, C2H4, C5H4, C4H2
% CH3 C2H2n+2, C5H4, C4H2 legend

%--------------------------------------------------------------------
\section{Methodology} \label{sec:method}

\subsection{System parameter and stellar input spectrum} \label{sec:star}
LHS~1140 is a close-by M4.5-type main-sequence red dwarf 14.993$\pm$0.015~pc away from the Earth \citep{gaia2018} with an effective temperature, T$_\text{eff}$, of 3219$\pm$39~K, a radius, $R$, of 0.2139$\pm$0.0041~$R_{\sun}$ and a mass, $M$, of 0.179$\pm$0.014~$M_{\sun}$ \citep{ment2019}. The star is known to host two rocky planets, LHS~1140~b and LHS~1140~c \citep{dittmann2017,ment2019}. In this study, we simulate the potential atmosphere of the habitable zone planet LHS~1140~b by using a radius of 1.727$\pm$0.032~$R_{\oplus}$, a mass of 6.98$\pm$0.89~$M_{\oplus}$ and a surface gravity, $g$, of 23.7$\pm$2.7 ms$^{-2}$ \citep{ment2019}. We do not expect that our results would change significantly when using the slightly lower planetary mass of 6.48$\pm$0.46~$M_{\oplus}$ suggested by \citet{lillo-box2020}.
% lower density compared to dittmann2017
The planet receives an incident flux of 0.503$\pm$0.03~$S_{\sun}$ and orbits its host star in $\sim$24.7~days. 

The stellar spectrum has not been measured for LHS~1140. However, the  FUV$_{1344-1786\AA}$/NUV$_{1711-2831\AA}$ ratio was determined to be 0.303$^{+0.090}_{-0.080}$  \citep{spinelli2019}. 
In the UV range up to 400~nm, we use the adapted panchromatic Spectral Energy Distribution (SED) of Proxima Centauri from the MUSCLES database version 22  \citep{france2016,loyd2016} with an FUV$_{1344-1786\AA}$/NUV$_{1711-2831\AA}$ ratio of 0.313. In the visible and NIR we take the SED from GJ~1214 with stellar parameters similar to LHS~1140 \citep[$T_\text{eff}$=3252$\pm$20~K, $R$=0.211$\pm$0.011~$R_{\sun}$, $M$=0.176$\pm$0.009~$M_{\sun}$,][]{anglada2013}.

\subsection{Model description and updates} \label{sec:model_updates}

%We use the steady-state, cloud-free, radiative-convective photochemistry model 1D-TERRA to simulate a range of potential atmospheres of LHS~1440~b. 
In this study, we use the radiative-convective photochemistry model 1D-TERRA.
The code dates back to early work by \citet{Kasting1986, pavlov2000} and \citet{segura2003} and has been considerably extended by e.g. \citet{grenfell2007,rauer2011,vonparis2011,grenfell2013,vonparis2015} and \citet{gebauer2017}.
%Earlier versions the model have been used to simulate the atmosphere of e.g. early Earth and Earth-like atmospheres around different types of main-sequence stars \citep[see e.g.][]{}. 
Recently a major update of the climate radiative transfer module \citep[called REDFOX;][]{scheucher2020} and chemistry module \citep[called BLACKWOLF;][]{wunderlich2020} enabled e.g. CO$_2$- and H$_2$-dominated atmospheres to be consistently simulated. 

REDFOX includes absorption of 20 molecules\footnote{CH$_3$Cl, CH$_4$, CO, CO$_2$, H$_2$, H$_2$O, HCl, HCN, HNO$_3$, HO$_2$, HOCl, N$_2$, N$_2$O, NH$_3$, NO, NO$_2$, O$_2$, O$_3$, OH, and SO$_2$} using spectroscopic cross sections from the HITRAN 2016 line list \citep{gordon2017} and 81 molecules using UV and VIS cross sections mainly taken from the MPI Mainz Spectral Atlas \citep{keller2013} as described in \citet{scheucher2020} and \citet{wunderlich2020}. 
Additionally Rayleigh scattering of eight molecules\footnote{CO, CO$_2$, H$_2$O, N$_2$, O$_2$, H$_2$, He, and CH$_4$}, Mlawer-Tobin-Clough-Kneizys-Davies absorption \citep[MT\_CKD;][]{mlawer2012} and CIAs of H$_2$-H$_2$, H$_2$-He, CO$_2$-H$_2$ CO$_2$-CH$_4$ and CO$_2$-CO$_2$ are considered \citep[see][for details]{scheucher2020}. 

The globally-averaged zenith angle is set to 60$^\circ$ in the climate module and 54.5$^\circ$ in the chemistry module in order to fit the  observed O$_3$ column of $\sim$300 Dobson Units (DU) on Earth \citep[see e.g.][]{degrandpre2000,thouret2006}. 
The atmosphere in the climate module is divided into 101 pressure levels and the chemistry module into 100 pressure layers. 
The eddy diffusion profile can be calculated according to the parameterization shown in \citet{wunderlich2020} or set to a given profile. Unless indicated otherwise, we use a parameterized eddy diffusion profile. The photochemical module accounts for dry and wet deposition, as well as surface emission fluxes and atmospheric escape \citep[see details in][]{wunderlich2020}. For wet deposition we use the parameterization of \citet{giorgi1985} and the tropospheric lightning emissions of nitrogen oxides, NO$_x$ (here defined as NO + NO$_2$) are based on the Earth lightning model of \citet{chameides1977}.

In the current paper, we additionally introduced some minor updates compared to the photochemical model described in \citet{wunderlich2020}. Recently, the water (H$_2$O) cross section between 186--230~nm has been measured by \citet{ranjan2020} for a temperature of 292~K. We use this new cross section data in the current study. 
%In \citet{wunderlich2020} we followed the recommendation by \citet{burkholder2015} and used the measurements from \citet{parkinson2003} up to 198~nm. This data set overestimates the water cross section compared to \citet{ranjan2020} between 192--198~nm by a factor up to 10. 
However, the weak NUV flux of M dwarfs suggests that the water photolysis is not affected significantly by the usage of the new measurements \citep[see e.g.][]{wunderlich2019}.

Recently, \citet{greaves2020a} found evidence of phosphine (PH$_3$) absorption in the atmosphere of Venus. The presence of detectable amounts of PH$_3$ is still debated in the literature \citep{snellen2020, thompson2020, encrenaz2020, villanueva2020, mogul2020,greaves2020b,greaves2020c} and chemical and biological processes, leading to its production, are not well known \citep{greaves2020a,bains2020,lingam2020}. However, \citet{sousa-silva2020} suggest that in H$_2$- and CO$_2$-dominated atmospheres chemical sinks of PH$_3$ are reduced compared to Earth, favoring a potential detection in such an environment. 
In the atmosphere of gas giants the thermodynamical formation of PH$_3$ is favored, where the pressure, temperature and the concentration of H$_2$ are sufficiently high \citep[see e.g.][]{visscher2006}.  
For rocky, potential habitable planets such conditions are not expected making PH$_3$ a reasonable candidate biosignature gas in a reduced atmosphere. 

The chemical network of BLACKWOLF, as presented in \citet{wunderlich2020}, did not include the chemical production and destruction of PH$_3$. Hence, we consider in the present work 16 additional phosphorous containing reactions (see Table~\ref{tab:reactions_ph3}). 
To calculate the wet deposition of PH$_3$ we use the Henry's Law constant from \citet{fu2013}. We do not consider any sink reaction for tetraphosphorus (P$_4$). At low temperatures P$_4$ is expected to sublimate without undergoing chemical reaction or photolysis \citep[see e.g.][]{kaye1984}. Hence, we assume that all P$_4$ is deposited or removed from the atmosphere in order to avoid a runaway effect.

With the new chemical reactions from Table~\ref{tab:reactions_ph3} we repeated the validation of modern Earth with 1D-TERRA shown in \citet{wunderlich2020}. The additional consideration of PH$_3$ has no significant impact on the concentration of key species in the atmosphere of modern Earth. 
PH$_3$ was measured locally on Earth with concentrations ranging between  1$\times$10$^{-15}$ (ppq) and 1$\times$10$^{-9}$ (ppb) \citep[see][and references therein]{pasek2014,bains2019,sousa-silva2020}. 
1D-TERRA suggests a global and annual mean surface mixing ratio of  1$\times$10$^{-12}$ (1~ppt) when using an assumed surface emission flux of 1$\times$10$^{8}$ molecules~cm$^{-2}$~s$^{-1}$. 

\begin{table}
\centering
\caption{Phosphorous containing reactions added to the photochemical reaction scheme.}
\label{tab:reactions_ph3}     
\centering                                     
\begin{tabular}{llc}        
\hline\hline 
    Reaction &  Reaction Coefficients & Ref. \\
    \hline 
PH$_3$+O$^1$D $\rightarrow$ PH$_2$+OH	&	$4.75\times10^{-11}$	&	(1)	\\
PH$_3$+OH $\rightarrow$ PH$_2$+H$_2$O	&	$2.71\times10^{-11} \cdot e^{-155/T}$	&	(2)	\\
PH$_3$+O $\rightarrow$ PH$_2$+OH	&	$9.95\times10^{-38}$ 	&	(3)	\\
PH$_3$+H $\rightarrow$ PH$_2$+H$_2$	&	$7.22\times10^{-11} \cdot e^{-886/T}$	&	(4)	\\
PH$_3$ + Cl $\rightarrow$ PH$_2$+HCl	&	$2.36\times10^{-10}$	&	(5)	\\
PH$_3$+N $\rightarrow$ PH$_2$+NH	&	$4.00\times10^{-14}$	&	(6)	\\
PH$_3$+NH$_2$ $\rightarrow$ PH$_2$+NH$_3$	&	$1.00\times10^{-12} \cdot e^{-928/T}$	&	(7)	\\
PH$_2$+H $\rightarrow$ PH+H$_2$	&	$6.20\times10^{-11} \cdot e^{-318/T}$	&	(8)	\\
H+PH $\rightarrow$ P+H$_2$	&	$1.50\times10^{-10} \cdot e^{-416/T}$	&	(8)	\\
P+PH $\rightarrow$ P$_2$+H	&	$5.00\times10^{-11} \cdot e^{-400/T}$	&	(8)	\\
PH$_2$+H+M $\rightarrow$ PH$_3$+M	&	$3.70\times10^{-10} \cdot e^{-340/T}$	&	(8)	\\
PH+H$_2$+M $\rightarrow$ PH$_3$+M	&		$3.00\times10^{-36} \cdot N$	&	(8)	\\
P+P+M $\rightarrow$ P$_2$+M	&	$1.40\times10^{-33} \cdot e^{500/T} \cdot N$	&	(8)	\\
P+H+M $\rightarrow$ PH+M	& $3.40\times10^{-33} \cdot e^{173/T} \cdot N$	&	(8)	\\
P$_2$+P$_2$+M $\rightarrow$ P$_4$+M	& $1.40\times10^{-33} \cdot e^{500/T} \cdot N$	&	(9)	\\
PH$_3$+hv $\rightarrow$ PH$_2$+H	&	see table notes	&	(10)	\\
    \hline                                            
\end{tabular}
\tablefoot{Bi-molecular reaction coefficients are shown in cm$^3$ s$^{-1}$ and termolecular reactions are in cm$^6$ s$^{-1}$. The unit of temperature, $T$, is K and the unit of number density, $N$, is cm$^{-3}$. All reactions are valid for temperatures around 298~K. Photolysis cross sections are taken from \citet{chen1991} between 120 and 230~nm. The quantum yield is assumed to be unity.}
\tablebib{(1)~\citet{nava1989}; (2)~\citet{fritz1982}; (3)~\citet{wang2005}; (4)~\citet{arthur1997}; (5)~\citet{iyer1983}; (6)~\citet{hamilton1985}; (7)~\citet{bosco1983}; (8)~\citet{kaye1984}; (9)~rate assumed to be the same as the reaction P+P+M$\rightarrow$ P$_2$+M; (10)~ \citet{chen1991} }
\end{table}

Additionally to the validation of 1D-TERRA against modern Earth, in \citet{scheucher2020} and \citet{wunderlich2020} the climate and chemistry  modules were validated against Mars and Venus-like conditions to show that the model is able to predict consistently N$_2$-O$_2$ and CO$_2$-dominated atmospheres.
In this work we validate the model against H$_2$-dominated and CH$_4$-rich atmospheres by simulating the atmosphere of Neptune in Appendix~\ref{app:neptune} and Titan in Appendix~\ref{app:Titan}.
 
\subsection{Climate-only runs} \label{sec:clima_scenario}
\begin{table}
\centering
\caption{Climate-only scenarios: surface pressure range, $p_0$ in bar, and mixing ratios, $f$, of N$_2$, CO$_2$, H$_2$ and He considered for LHS~1140~b.}
\label{tab:composition_clima}     
\centering                                     
\begin{tabular}{cccccc}        
\hline\hline 
    Scenario & p$_0$ & N$_2$ & CO$_2$ &  H$_2$ & He  \\
    \hline
    I & 0.7--100 & 0.9996 & 4$\times$10$^{-4}$ & 0 & 0 \\
    II & 0.1--22 & 0 & 1 & 0 & 0 \\
    III & 0.1--6 & 0 & 0 & 0.8 & 0.2 \\
    \hline 
    \hline
\end{tabular}
\end{table}

We perform climate-only runs of N$_2$-dominated, CO$_2$, and H$_2$-He atmospheres with the radiative transfer module REDFOX and vary the surface pressures in order to investigate for which atmospheric conditions LHS~1140~b could be habitable at the surface (see Table~\ref{tab:composition_clima}). The mixing ratios of the species are constant over height. We consider pressures leading to surface temperatures between 220~K  \citep[approximated limit of open water with ocean heat transport in climates of tidally locked exoplanets around M dwarf stars, see][]{HuYang2014,Checlair2017,Checlair2019} and 395~K \citep[see][]{Clarke2004,McKay2014}. Further we limit our calculation to 100~bars surface pressure since massive envelopes are not expected due to the high bulk density of the planet \citep{ment2019}. 

We include absorption by the major radiative species \citep{scheucher2020}. For the H$_2$O profile we use a constant relative humidity of 80\% up to the tropopause. Above the tropopause the H$_2$O profile is set to a constant abundance based on its value at the cold trap.
For the N$_2$ atmospheres we assume an Earth-like CO$_2$ level of 400~ppm \citep[see e.g.][]{monastersky2013} and for the H$_2$-dominated atmospheres we use 80\% H$_2$ and 20\% He.

\subsection{Coupled Climate-Chemistry runs} \label{sec:scenarios}
Here we apply the coupled version of 1D-TERRA to simulate the potential atmospheric temperature and composition profiles of LHS~1140~b.
%In the REDFOX radiative transfer module we use MT\_CKD absorption of H$_2$O and CO$_2$ and CIA of H$_2$-H$_2$, H$_2$-He, CO$_2$-H$_2$ \citep{wordsworth2017}, CO$_2$-CH$_4$ \citep{wordsworth2017} and CO$_2$-CO$_2$ \citep{gruszka1997,baranov2004}.
All simulations assume a constant relative humidity of 80\% from the surface to the cold trap. The surface albedo is set to 0.255, which is the value needed to achieve a mean surface temperature of 288.15~K for the Earth around the Sun \citep[see][]{scheucher2020,wunderlich2020}. 

%\subsubsection{Model scenarios} 
%thin h2 atmosphere: https://static.exo3online.de/userdata/sdfjellsmgfletrm/events/ibmfjyvkmmpvcunn/ibmfjyvkmmpvcunn_presentation.pdf

\begin{table}
\centering
\caption{Clima-chemistry scenarios: surface mixing ratio of CO$_2$, H$_2$, He, and CH$_4$.}
\label{tab:composition}     
\centering                                     
\begin{tabular}{ccccc}        
\hline\hline 
    Scenario & CO$_2$ &  H$_2$ & He  & CH$_4$ \\
    \hline 
    1a & \multirow{ 3}{*}{1$\times$10$^{-9}$}	&	\multirow{ 3}{*}{fill gas}	&	\multirow{ 3}{*}{0.2} & 1$\times$10$^{-6}$ \\
    1b & & & &	1$\times$10$^{-3}$  \\
    1c & & & &	$0.03$ \\
    \hline 
    2a & \multirow{ 3}{*}{1$\times$10$^{-3}$}	&	\multirow{ 3}{*}{fill gas}	&	\multirow{ 3}{*}{0.1998}	& 1$\times$10$^{-6}$\\
    2b & & & &	 1$\times$10$^{-3}$   \\
    2c & & & &	$0.03$ \\
    \hline 
    3a & \multirow{ 3}{*}{0.01}	&	\multirow{ 3}{*}{fill gas}	&	\multirow{ 3}{*}{0.198}	& 1$\times$10$^{-6}$\\
    3b & & & &	 1$\times$10$^{-3}$   \\
    3c & & & &	$0.03$ \\
    \hline 
    4a & \multirow{ 3}{*}{0.1}	&	\multirow{ 3}{*}{fill gas}	&	\multirow{ 3}{*}{0.18} & 1$\times$10$^{-6}$	\\
    4b & & & &	 1$\times$10$^{-3}$   \\
    4c & & & &	$0.03$ \\
    \hline 
    5a & \multirow{ 3}{*}{0.3}	&	\multirow{ 3}{*}{fill gas}	&	\multirow{ 3}{*}{0.14} & 1$\times$10$^{-6}$	\\
    5b & & & &	 1$\times$10$^{-3}$   \\
    5c & & & &	$0.03$ \\
    \hline 
    6a & \multirow{ 3}{*}{fill gas}	&	\multirow{ 3}{*}{0.4}	&	\multirow{ 3}{*}{0.1} &	1$\times$10$^{-6}$ \\
    6b & & & &	 1$\times$10$^{-3}$   \\
    6c & & & &	$0.03$ \\
    \hline 
    7a & \multirow{ 3}{*}{fill gas}	&	\multirow{ 3}{*}{0.24}	&	\multirow{ 3}{*}{0.06} & 1$\times$10$^{-6}$	\\
    7b & & & &	  1$\times$10$^{-3}$   \\
    7c & & & &	$0.03$ \\
    \hline 
    8a & \multirow{ 3}{*}{fill gas}	&	\multirow{ 3}{*}{0.08}	&	\multirow{ 3}{*}{0.02} & 1$\times$10$^{-6}$	\\
    8b & & & &	  1$\times$10$^{-3}$   \\
    8c & & & &	$0.03$ \\
    \hline 
    9a & \multirow{ 3}{*}{fill gas}	&	\multirow{ 3}{*}{8$\times$10$^{-3}$}	&	\multirow{ 3}{*}{2$\times$10$^{-3}$} & 1$\times$10$^{-6}$	\\
    9b & & & &	  1$\times$10$^{-3}$   \\
    9c & & & &	$0.03$ \\
    \hline 
    10a & \multirow{ 3}{*}{fill gas}	&	\multirow{ 3}{*}{8$\times$10$^{-6}$}	&	\multirow{ 3}{*}{2$\times$10$^{-6}$} & 1$\times$10$^{-6}$	\\
    10b & & & &	  1$\times$10$^{-3}$ \\
    10c & & & &	$0.03$ \\
    \hline 
    \hline                                            
\end{tabular}
\tablefoot{Fill gas denotes the main constituent of the atmosphere. Scenario 1c has a composition similar to Neptune: a H$_2$-dominated atmosphere with 20\% He \citep{williams2004}, 1~ppb CO$_2$ \citep{meadows2008} and 3\% CH$_4$ \citep{irwin2019}. Scenario 10a has a CO$_2$-dominated atmosphere with $\sim$10~ppm H$_2$ and He similar to Mars and Venus \citep{krasnopolsky2005,krasnopolsky2001}. For each of the ten main scenarios regarding the main composition of the atmosphere we consider three different boundary conditions of CH$_4$ (see text). 
%(a) surface flux only from geological sources \citep{catling2017}; (b) flux from pre-industrial biomass emissions and geological sources \citep{etiope2009} and (c) a fixed surface mixing ratio of 3\%.
}
\end{table}

\begin{table}
\centering
\caption{Assumed surface emissions (in molecules cm$^{-2}$ s$^{-1}$) and dry deposition velocities (in cm s$^{-1}$) for model runs. }
\label{tab:Earth_flux}     
\centering                                     
\begin{tabular}{lcc}        
\hline\hline 
    Species &  Emissions (molec. cm$^{-2}$ s$^{-1}$) & $\nu_{\text{dep}}$ (cm s$^{-1}$).    \\
    \hline 
  O$_2$      & 1.21$\times$10$^{12}$  & $1\times10^{-8}$ \\
  CH$_3$Cl   &  1.39$\times$10$^{10}$  & $1\times10^{-8}$\\  
  PH$_3$   &  1.00$\times$10$^{10}$  &  $1\times10^{-8}$\\  
  NH$_3$     & 8.38$\times$10$^{10}$  & $1\times10^{-8}$\\
  N$_2$O     & 7.80$\times$10$^{8}$   &  $1\times10^{-8}$ \\
  CO         & 1.07$\times$10$^{11}$  & 0.02\\
  NO         & 3.38$\times$10$^{8}$   &  0.02\\
  H$_2$S     & 3.73$\times$10$^{9}$   &  0.02\\
  SO$_2$     &  1.34$\times$10$^{10}$  & 0.02\\
  OCS        &  1.42$\times$10$^{8}$  & 0.02\\
  HCN        &  1.27$\times$10$^{7}$  & 0.02\\
  CH$_3$OH   &  3.35$\times$10$^{10}$ & 0.02\\
  CS$_2$     &  5.05$\times$10$^{8}$  & 0.02\\  
  C$_2$H$_6$ &  8.55$\times$10$^{8}$  & 0.02 \\
  C$_3$H$_8$ &  9.51$\times$10$^{8}$  & 0.02 \\
  HCl        &  5.57$\times$10$^{9}$  & 0.02\\
    \hline                                            
\end{tabular}
\end{table}

Table \ref{tab:composition} shows the 30 scenarios performed with the coupled-climate model. All scenarios assume a constant surface pressure of 2.416 bar, corresponding to the atmospheric mass of the Earth assuming a surface gravity g of 23.7 ms$^{-2}$ for LHS 1140 b \citep{ment2019}. We chose this moderate surface pressure for the following reasons: LHS~1140~b has a high bulk density, $\rho$, of 7.5$\pm$1.0 cm$^{-3}$ \citep{ment2019}. Hence, it is unlikely that the planet has a thick H$_2$ or He envelope. However, the enhanced gravity compared to Earth results in reduced H$_2$ escape rates \citep[see e.g.][]{pierrehumbert2011}. This is supported by theoretical studies showing that cool and/or massive Super-Earths can retain small residual H$_2$/He envelopes at the end of the core-powered mass loss \citep{misener2020,gupta2019,ginzburg2016}. The secondary outgassing of CO$_2$ is expected to be small for a Super-Earth like LHS 1140 b with a mass of $\sim$7~$M_{\oplus}$ \citep[see e.g.][]{dorn2018,noack2017}. Hence, we do not consider thick CO$_2$ atmospheres as on Venus.

In this study we vary atmospheric mixtures of H$_2$-He and CO$_2$ in 10 steps (see Table \ref{tab:composition}).
For each of the steps we consider in addition three different boundary conditions for CH$_4$. 
The CH$_4$ abundance can have a large impact on surface temperature and habitability \citep[see e.g.][]{pavlov2000,ramirez2018}. Also the detectability of atmospheric spectral features on exoplanets can largely depend on the CH$_4$ inventory due to haze formation or CH$_4$ absorption \citep[see e.g.][]{arney2016,arney2017,lavvas2019,wunderlich2019}.
%The molecular fraction of CH$_4$ in the solar system varies from ppb on Mars \citep{webster2018} and ppm on Earth to several percent on Neptune and Titan \citep{irwin2019,niemann2010}.
We vary the boundary conditions of CH$_4$ as follows:

%noack 2017 and dorn 2018: outgassing just up to 6 mass
%increase of CO2 due to haze -> CO2/CH4 ratio
\begin{enumerate}[a:]
    \item The "low CH$_4$" scenarios assume the volume mixing ratio (vmr) of CH$_4$ to be constant at 1$\times$10$^{-6}$ at the surface, corresponding roughly to the surface CH$_4$ concentration in the pre-industrial era on Earth \citep[see e.g.][]{etheridge1998}. 
    \item The "medium CH$_4$" scenarios use a constant CH$_4$ vmr of 1$\times$10$^{-3}$ at the surface. Model studies such as \citet{rugheimer2015} and \citet{wunderlich2019} suggest a CH$_4$ vmr of roughly 1$\times$10$^{-3}$ for Earth-like planets in the HZ around mid-M dwarfs using a surface emission of 1.4$\times$10$^{11}$~molec.~cm$^{-2}$~s$^{-1}$ as measured on Earth \citep[e.g.][]{lelieveld1998}.  
    \item The "high CH$_4$" scenarios assume that the surface vmr of CH$_4$ is constant at 3\%. This is consistent with the observed main composition of the lower atmosphere of Neptune with 2--4\% CH$_4$ at 2~bar \citep[see e.g.][]{irwin2019}.
\end{enumerate}

%We vary the main composition of the atmosphere in 10 steps from a H$_2$-He atmosphere to a CO$_2$-dominated atmosphere (see Table~\ref{tab:composition}). The surface pressure is assumed to be constant at 2.4~bar, corresponding to one Earth atmospheric mass on LHS~1140~b with a surface gravity $g$ of 23.7 ms$^{-2}$ \citep{ment2019}. 
%This surface pressure leads to habitable surface conditions for H$_2$ and CO$_2$ dominated atmospheres (see Section~\ref{sec:habitability}). 
We assume the same boundaries for all 30 scenarios (except for CO$_2$, H$_2$, He, and CH$_4$). For O$_2$, CO, H$_2$S, NO, N$_2$O, OCS, HCN, CS$_2$, CH$_3$OH, C$_2$H$_6$, C$_3$H$_8$ and HCl we assume pre-industrial Earth-like biogenic fluxes (see Table~\ref{tab:Earth_flux}). 
Regarding NH$_3$, PH$_3$ and CH$_3$Cl we use larger biogenic fluxes than measured on Earth, assuming that an H$_2$-dominated atmosphere could favor the biogenic production of these species, since e.g. H$_2$ can act as a nutrient.
We assume a biogenic NH$_3$ flux of 8.38$\times$10$^{10}$ molecules cm$^{-2}$ s$^{-1}$ corresponding to mean global emissions on a hypothetical cold Haber world \citep{seager2013}. This NH$_3$ flux is $\sim$100 times larger than observed on pre-industrial Earth \citep{bouwman1997}. The assumed biogenic emissions of CH$_3$Cl are assumed to be 100 times larger than on pre-industrial Earth \citep{seinfeld2016}. The biogenic surface flux of PH$_3$ is taken from \citet{sousa-silva2020}.
Additionally, we apply biogenic and volcanic emissions as measured on Earth \citep[see Table~\ref{tab:Earth_flux} and][for references]{wunderlich2020}.
We assume a non-zero dry deposition velocity, $\nu_{\text{dep}}$, for all species to reduce a potential runaway effect \citep[see e.g.][]{hu2020}. 
For CH$_4$, O$_2$, CH$_3$Cl, PH$_3$, NH$_3$ and N$_2$O we assume a $\nu_{\text{dep}}$ of $1\times10^{-8}$~cm~s$^{-1}$. For O$_2$ this value was used by other model studies \citep[e.g.][]{arney2016,hu2020,wunderlich2020} and provides an upper estimation of how much of these gases could be accumulated for the scenarios assumed.
For all other species we use a $\nu_{\text{dep}}$ of 0.02~cm~s$^{-1}$, following \citet{hu2012} and \citet{zahnle2008}.

\subsection{Transmission} \label{sec:spectra}
%and emission spectra
The simulated atmospheres serve as input to compute transmission %and emission
spectra with the "Generic Atmospheric Radiation Line-by-line Infrared Code" \citep[GARLIC;][]{schreier2014,schreier2018}. Line parameters and CIAs are taken from the HITRAN database \citep{gordon2017,karman2019}. We consider the CKD continuum model for H$_2$O \citep{clough1989} and  Rayleigh extinction for H$_2$, He, CO$_2$, H$_2$O, N$_2$, CH$_4$, CO, N$_2$O and O$_2$ \citep{murphy1977, sneep2005, marcq2011}. In the visible we use the cross sections at room temperature (298~K) listed in Table 3 of \citet{wunderlich2020}. 

We assume cloud-free conditions for all simulated transmission spectra.
%and emission spectra. 
We do, however, consider extinction from uniformly distributed aerosols with an optical depth, $\tau_\text{A}$, at wavelength, $\lambda$ ($\mu$m), following \citet{angstrom1930,allen1976} and \citet{yan2015}: 

\begin{equation} \label{eq:sigma_aer} 
  \tau_{\text{A}} = \beta \cdot N_\text{c} \cdot \lambda^{-\alpha},
\end{equation}
\noindent
with the column density, $N_\text{c}$, in molecules~cm$^{-2}$.
For clear sky conditions with weak scattering by haze or dust we set $\alpha$ to 1.3, representing an average measured value on Earth following the Junge distribution \citep[see e.g.][]{angstrom1961} and we set the coefficient $\beta$ to $1.4\times10^{-27}$ following \citet{allen1976}.
For hazy conditions we assume $\alpha$ to be 2.6 and set $\beta$ to $6.0\times10^{-25}$, representing the best fit to extinction by hazes on Titan (see Appendix~\ref{app:hazes}). 
Note, that we do not consider the production of hazes in H$_2$- and CO$_2$-dominated atmospheres in our model. The assumed impact of hazes on the spectral appearance of the simulated atmospheres should therefore only serve as a rough estimation and further investigation is needed to test the validity of this assumption.

We express the transmission spectra as wavelength dependent atmospheric transit depth, which contains the contribution of the atmosphere to the total transit depth without the contribution from the solid body \citep[see more details in][]{schreier2018,wunderlich2020}. 
%The emission spectra are calculated assuming a down-looking observer at a viewing angle of 38$^{\circ}$, which accounts for the average stellar illumination \citep[this angle was also used by e.g.][]{segura2003,rauer2011,katyal2019}.

\subsection{Signal-to-noise ratio (S/N)} \label{sec:snr}

\begin{table}
\centering
\caption{Wavelength coverage and resolving power, $R$, of the instruments on JWST and ELT used to calculate the number of transits required to detect spectral features in the atmosphere of LHS~1140~b. }
\label{tab:instruments}  
\centering                                     
\begin{tabular}{lccc}        
\hline\hline 
    Telescope - Instrument & Wavelength & $R$ & Ref.  \\
    \hline 
  JWST - NIRSpec PRISM    & 0.6--5.3~$\mu$m  & $\sim$100  &  (1)    \\
  JWST - MIRI LRS     & 5.0--12~$\mu$m  & $\sim$100  & (2)     \\
  ELT - HIRES    & 0.37--2.5~$\mu$m  & 100,000  & (3)     \\
  ELT - METIS (HRS)  & 2.9--5.3~$\mu$m  & 100,000  & (4)     \\
    \hline  
\end{tabular}
\tablebib{(1) \citet{birkmann2016b}; (2) \citet{kendrew2015}; (3) \citet{marconi2016}; (4) \citet{brandl2016} }
\end{table}

An important aim of this study is to determine whether molecular absorption features are detectable with the JWST and ELT (see Table~\ref{tab:instruments}). 
For low resolution spectroscopy (LRS) we calculate the required number of transits with JWST NIRSpec PRISM and MIRI LRS with the method presented in \citet{wunderlich2019}. LHS~1140 exceeds the brightness limits of NIRSpec PRISM leading to a saturation of the detector between 1 and 2~$\mu$m. Hence, we consider partial saturation of the detector as proposed by \citet{batalha2018} and exclude the wavelength range between 1 and 2~$\mu$m from our analysis. The results do not significantly differ when we use medium resolution filters such as NIRSpec G235M and G395M \citep{birkmann2016b}, which do not saturate for LHS~1140. However, both filters cannot be used simultaneously which would decrease the wavelength coverage compared to NIRSpec PRISM and the additional binning of the spectral data with larger resolving power might lead to enhanced red noise.

With high resolution spectroscopy (HRS) we investigate the potential detection of spectral lines from NH$_3$, PH$_3$, CH$_3$Cl and N$_2$O with the cross-correlation technique using the HIRES and METIS instruments on ELT. We use the same approach to estimate the number of transits which are necessary to detect the molecules with the cross-correlation method as described in \citet{wunderlich2020}. 
The signal to noise of the star per transit is calculated with the European Southern Observatory Exposure Time Calculator\footnote{\url{https://www.eso.org/observing/etc/bin/gen/form?INS.NAME=ELT+INS.MODE=swspectr}} (ESO ETC) Version 6.4.0 from November 2019 \citep[see updated documentation\footnote{\url{https://www.eso.org/observing/etc/doc/elt/etc_spec_model.pdf}} from][]{liske2008}. As input for the ETC we use the stellar spectrum described in Sec.~\ref{sec:star} and scale it to the K-band magnitude of 8.821 \citep{cutri2003} in order to obtain the input flux distribution. 
Equivalent to \citet{wunderlich2020} we use a mean throughput of 10\% for ELT HIRES and METIS. Previous studies investigating the feasibility of detection of e.g. O$_2$ in Earth-like atmospheres assumed a more optimistic throughput of 20\% \citep{snellen2013,rodler2014,serindag2019}.  The sky conditions are set to a constant airmass of 1.5 and a precipitable water vapour (PWV) of 2.5. For each wavelength band we change the radius of the diffraction limited core of the point spread function according to the recommendation in \citet{liske2008}.

\section{Results and Discussion} \label{sec:results}
\subsection{Surface habitability} \label{sec:habitability}

%https://www.pnas.org/content/111/2/629.short
%https://iopscience.iop.org/article/10.3847/1538-4357/aa80e1/pdf
%https://www.pnas.org/content/pnas/111/2/629.full.pdf
%https://iopscience.iop.org/article/10.1088/0004-637X/778/2/109/pdf

\begin{figure}
\centering
   \includegraphics[width=0.44\textwidth]{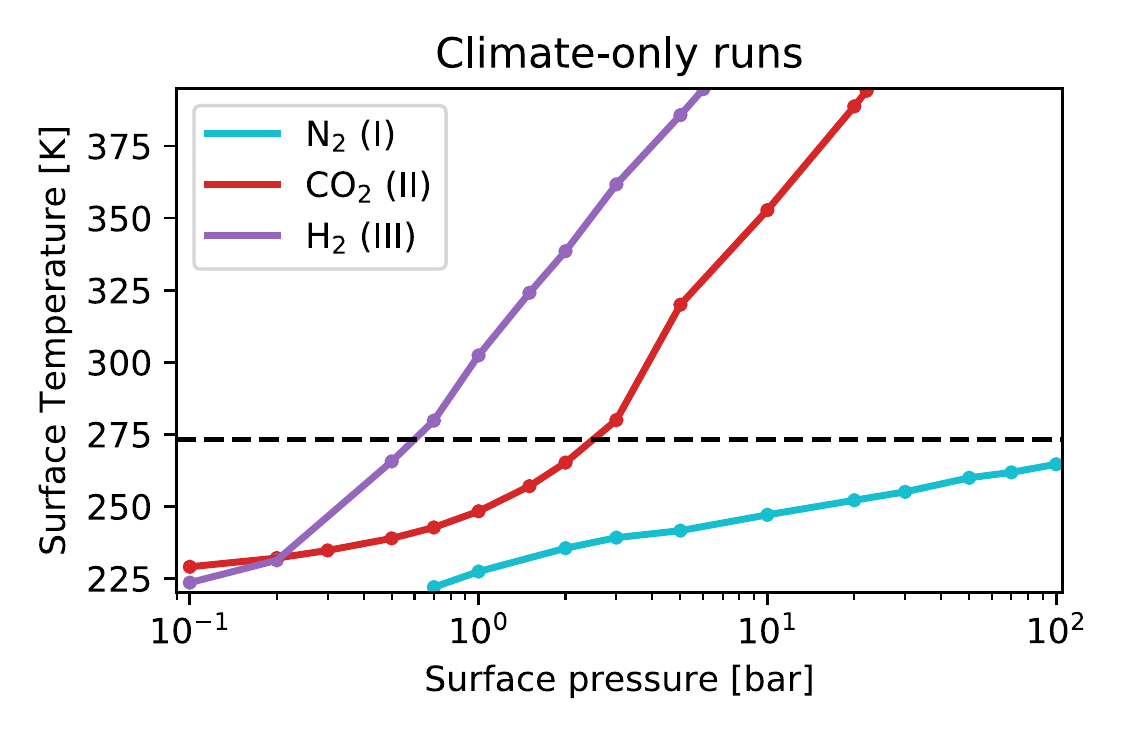}
     \caption{Surface temperature in Kelvin against surface pressure in bar for simulated climate-only N$_2$-, CO$_2$-, and H$_2$-dominated atmospheres of LHS~1140~b (see Table~\ref{tab:composition_clima}). Horizontal dashed line indicates 273~K. }
     \label{fig:tsurf}
\end{figure}

Figure~\ref{fig:tsurf} shows the surface temperatures of LHS~1140~b for N$_2$-, CO$_2$-, and H$_2$-dominated atmospheres with varying surface pressures, for the climate-only runs (without coupling to the photochemistry module, see Sec.~\ref{sec:clima_scenario}). Results suggest that thick N$_2$-dominated atmosphere on LHS-1140~b or substantial amounts of greenhouse gases such as CO$_2$ would be required to reach habitable surface temperatures. The simulations show that a CO$_2$ atmosphere requires a surface pressure of $\sim$2.5~bar to reach a global mean surface temperature above 273~K. This is comparable to the results of \citet{morley2017} who found that a 2~bar Venus-like atmosphere on LHS~1140~b would lead to a surface temperature of around 280~K. 

For an H$_2$ atmosphere CIA by H$_2$-H$_2$ extends the outer edge of the HZ \citep[see e.g.][]{pierrehumbert2011} compared to CO$_2$ atmospheres. The simulated H$_2$ atmospheres have up to 80~K warmer surface temperatures than the CO$_2$ atmospheres. Our results suggest that a surface pressure of 0.6~bar leads to a surface temperature of about 273~K on LHS~1140~b.

%- climate- chemistry model
%- stellar input
%- snr model

%- why 1e-2: co2 amount for outgassing like on Earth
%- why 3: lowest co2 where all scenario have habitable surface temperature
%--> 5????

\subsection{Atmospheric profiles} \label{sec:profiles}

\begin{figure*}
\centering
   \includegraphics[width=1.0\textwidth]{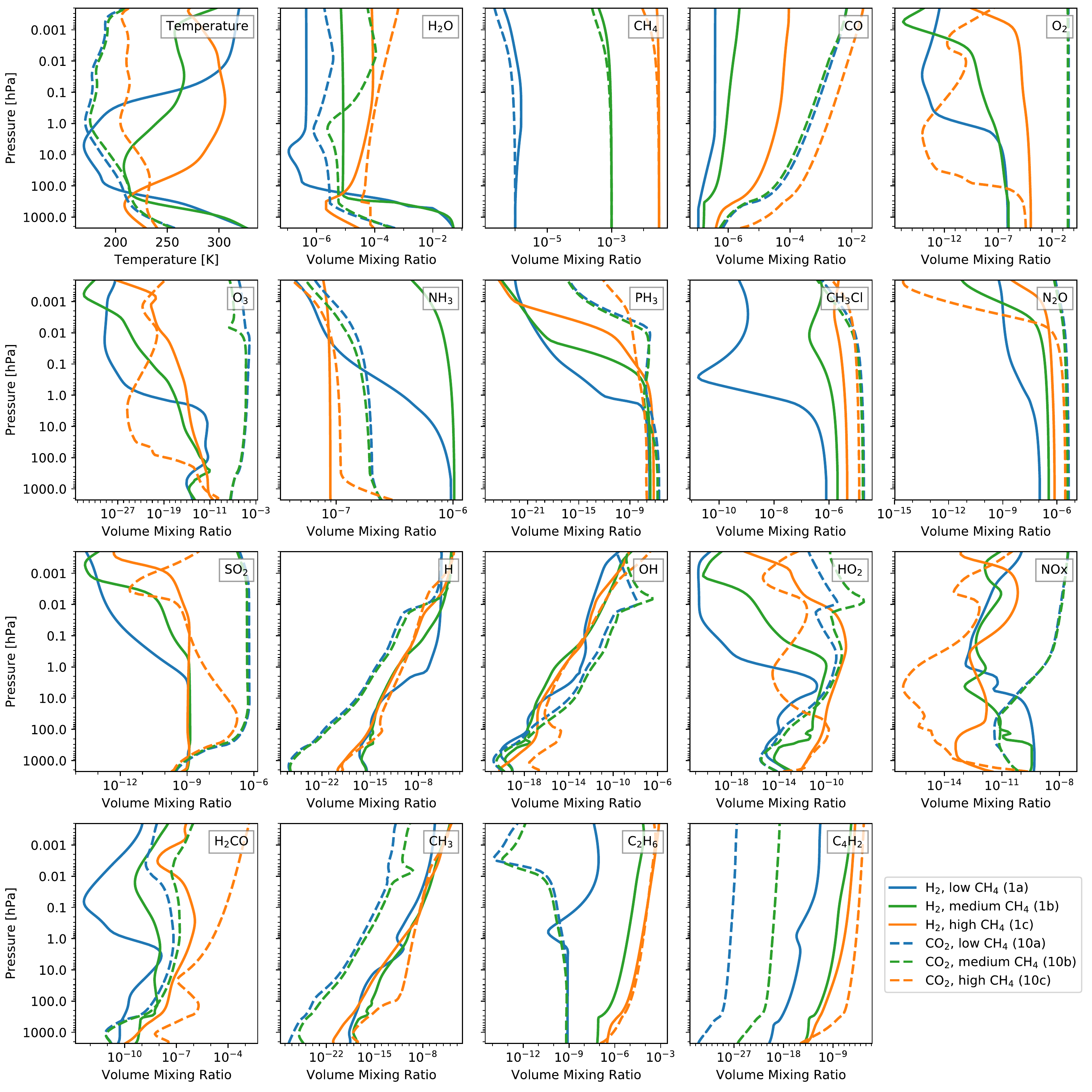}
     \caption{Simulated temperature and composition profiles of selected species of LHS~1140~b. Different colors represent the three types of scenarios considered: blue for low CH$_4$, green for medium CH$_4$ and orange for high CH$_4$. Solid lines represent H$_2$-dominated atmospheres (scenarios 1a, 1b, and 1c) whereas dashed lines show CO$_2$-dominated atmospheres (scenarios 10a, 10b, 10c). 
     %Note that the mixing ratios of C$_4$H$_2$ for scenarios 10a and 10b are below $10^{-20}$.
     }
     \label{fig:profiles}
\end{figure*}

\begin{figure}
\centering
   \includegraphics[width=0.49\textwidth]{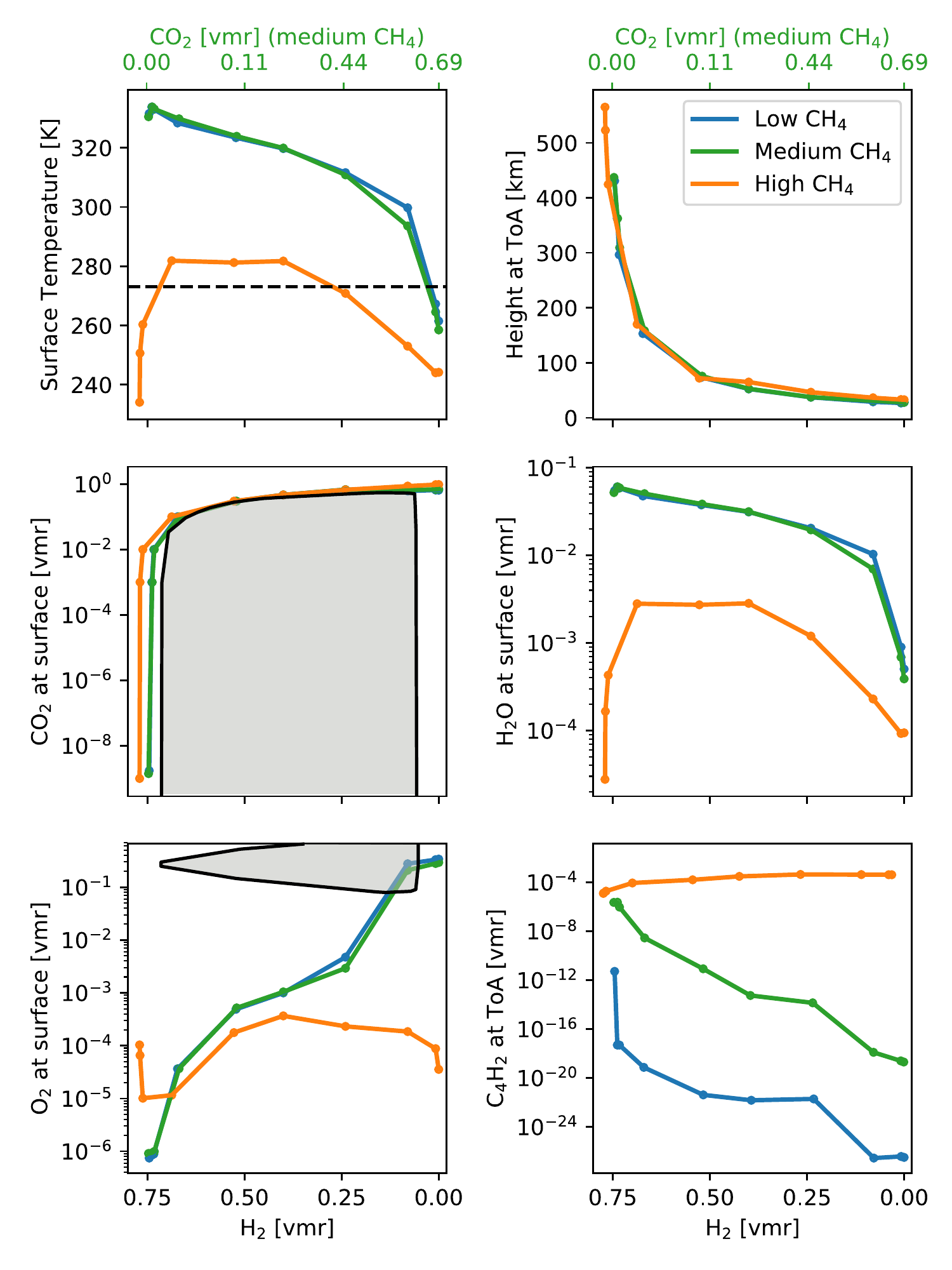}
     \caption{Change in the surface temperature in K, the atmospheric height at ToA ($\sim$0.01~Pa) in km and the surface volume mixing ratio (vmr) of H$_2$O, O$_2$, CO and CH$_4$ with decreasing concentrations of H$_2$ for scenarios with low (blue lines), medium (green lines) and high CH$_4$ scenarios. Note that the concentrations of CO$_2$ corresponding to H$_2$ vmrs of 0.75, 0.5, 0.25 and 0.0 are shown in green x-labels for the medium CH$_4$ scenarios. The colored dots represent the 10 individual main scenarios described in Table~\ref{tab:composition}. The dashed black line shows T=273~K. The gray shaded region represents the limits of combustion for H$_2$-CO$_2$ and H$_2$-O$_2$ gas mixtures \citep[see][]{grenfell2018}. Note that the assumed surface mixing ratios of H$_2$ and CO$_2$ for the main scenarios 5--8 are at the edge of the H$_2$-CO$_2$ combustion limit.}
     \label{fig:evol_prof}
\end{figure}

To simulate the atmospheric scenarios as described in Section \ref{sec:scenarios} we use the coupled version of 1D-TERRA.
Figure~\ref{fig:profiles} shows the temperature and composition profiles of selected species for H$_2$-dominated atmospheres with low concentrations of CO$_2$ (scenarios 1a, 1b, and 1c; see Sect.~\ref{sec:scenarios} and Table~\ref{tab:composition}) and CO$_2$-dominated atmospheres with low concentrations of H$_2$ (scenarios 10a, 10b, and 10c). 
Figure~\ref{fig:evol_prof} shows the surface temperature, the atmospheric height at the Top of Atmosphere (ToA) at $\sim$0.01~Pa, the surface vmr of CO$_2$, H$_2$O, and O$_2$, and the CH$_4$ vmr at the ToA with decreasing concentrations of H$_2$ (corresponding to increasing concentrations of CO$_2$, see Table~\ref{tab:composition}). 

\subsubsection{Temperature}\label{sec:profiles_t}
Figure~\ref{fig:profiles} suggests that the H$_2$-dominated atmospheres with low and medium CH$_4$ concentrations (corresponding to scenarios 1a and 1b respectively) show a similar temperature profile from the surface up to $\sim$200~hPa  with a strong warming towards the ToA.  
%The low CH$_4$ scenario has a lower temperature between 0.1--100~hPa due to weaker warming due to CH$_4$ absorption. At pressures less than 0.1~hPa a high concentration of CH$_4$ cools the atmosphere, leading to lower temperatures for the scenario 1b compared to scenario 1a at these altitudes.

The H$_2$-dominated atmosphere with a high CH$_4$ concentration (scenario 1c) results in a warm stratosphere due to CH$_4$ short-wave absorption and accordingly a cool troposphere, similar to the shape of the temperature profile behaviour on e.g. Titan \citep[see e.g.][]{fulchignoni2005,serigano2016}. The large concentration of CH$_4$ absorbs most of the stellar energy in the stratosphere, leading to reduced stellar irradiation reaching the troposphere \citep[see also][]{ramirez2018}. Note that 1D-TERRA does not consider the effect of hazes, which might be formed in such an environment \citep[see e.g.][]{he2018,horst2018}.

%This suggests that most of the stellar irradiation is absorbed in the stratosphere, resulting in a smaller increase of the surface temperature due to reduced stellar irradiation reaching the troposphere (see also Grenfell et al. 2013).

In CO$_2$-dominated atmospheres we simulate similar temperature profiles for low CH$_4$ concentrations (scenario 10a) compared to medium CH$_4$ concentrations (scenario 10b). There is no temperature inversion from O$_3$ absorption in the middle atmosphere due to the weak UV emission of M dwarfs \citep[see also e.g.][]{segura2005,grenfell2014,wunderlich2019,wunderlich2020}. 
The CO$_2$-dominated atmosphere with high abundances of CH$_4$ (scenario 10c) shows weak temperature variations in the range of $\sim$40~K through the simulated atmospheric profile. The high concentration of CH$_4$ leads to a warming of the atmosphere compared to the runs with lower CH$_4$ abundances, except near the surface, where the anti-greenhouse effect cools the atmosphere.

% evolution 
In Fig~\ref{fig:evol_prof} the low and medium CH$_4$ scenarios (blue and green lines, respectively) feature similar responses in surface temperatures with decreasing H$_2$.
%An increase of the CO$_2$ abundances from 1~ppb to 1\% in an H$_2$-dominated atmosphere has only a weak impact on the resulting surface temperature. A further increase of CO$_2$ leads to a decrease of the surface temperature by up to 70~K (see also Section~\ref{sec:habitability}).
The high CH$_4$ scenarios (orange line) first show a warming effect of about 50~K due to the decreased partial pressure of CH$_4$ on increasing the molecular weight towards CO$_2$-rich atmospheres (scenarios 1c to 4c). Note that the surface mixing ratio of CH$_4$ is kept constant at 3\% for all the high CH$_4$ scenarios. For CO$_2$-dominated atmospheres the surface temperature decreases when reducing the abundances of H$_2$ (scenarios 6c to 10c) due to the weaker warming from the H$_2$-H$_2$ CIA.

%shows up to 100~K lower surface temperatures than the other two scenarios due to strong absorption of stellar light by CH$_4$ in the stratosphere leading to an anti-greenhouse effect. For the run with 10\% CO$_2$ the surface temperature is increased by 50~K compared to the run with the 1~ppb of CO$_2$. 
%Note that we keep the mixing ratio of CH$_4$ constant at 3\% for all the runs of the high CH$_4$ scenario. Further, the surface pressure is kept constant at 2.416~bar for all the runs. Due to the larger molecular weight of CO$_2$-rich atmospheres, the partial pressure of CH$_4$ is decreased. This leads to lower absorption by CH$_4$, a cooling in the stratosphere and warming in the troposphere compared to the CO$_2$-poor runs. 

\subsubsection{H$_2$O}
The water profile in the lower atmosphere depends mainly on the assumed relative humidity and the temperature profile near the surface. For all simulations we assume the relative humidity to be constant at 80\% up to the tropopause. In the middle and upper atmosphere the H$_2$O profile is determined mainly by chemical production or loss and to a lesser extent by eddy mixing. 
H$_2$O is mainly destroyed by photolysis at wavelengths below 200~nm forming the hydroxyl radical (OH) and atomic hydrogen (H):
\begin{align}
\text{H$_2$O + h}\nu & \rightarrow \text{H + OH.} \tag{R1} \label{eq:H2O_hv}
\end{align}

The H$_2$-dominated atmospheres show weaker FUV absorption compared to the CO$_2$-dominated atmospheres. This weaker shielding effect leads to enhanced photolysis and less stratospheric water content for the scenarios with H$_2$-dominated atmospheres (Fig.~\ref{fig:profiles}). For the high CH$_4$ scenarios the water photolysis is weak due to strong FUV absorption from CH$_4$. In CO$_2$-dominated atmospheres H$_2$O is significantly reformed via HO$_x$-driven (HO$_x$ = H + OH + HO$_2$) oxidation of CH$_4$ at pressures less than 0.1 hPa \citep[see also][]{segura2005,grenfell2013,rugheimer2015,wunderlich2019}.

\subsubsection{O$_2$} \label{sec:profiles_O2}
All runs assume an Earth-like surface O$_2$ flux from photosynthesis (see Table~\ref{tab:Earth_flux}). However, in H$_2$-dominated atmospheres the oxygen content near the surface is only $\sim$1~ppm for low and medium CH$_4$ scenarios (blue and green solid line in Fig.~\ref{fig:profiles}). 
\citet{grenfell2018} suggest that the catalytic cycles including HO$_x$ and NO$_x$ leads to oxidation of H$_2$ into water. O$_2$ is mainly destroyed via photolysis or the three body reaction with atomic hydrogen:
\begin{align}
\text{O$_2$ + h}\nu & \rightarrow \text{O + O} \tag{R2} \label{eq:O2_hv} \\
\text{H + O}_2 \text{+ M} & \rightarrow \text{HO$_2$ + M} \tag{R3} \label{eq:H_O2}
\end{align}
The atomic hydrogen required for reaction~(\ref{eq:H_O2}) is mainly produced via:
\begin{align}
\text{OH + H}_2 & \rightarrow \text{H$_2$O + H.} \tag{R4} \label{eq:OH_H2}
\end{align}
OH can be formed via the reactions:
\begin{align}
\text{H + HO}_2 & \rightarrow \text{OH + OH} \tag{R5} \label{eq:H_HO2}\\
\text{H + NO}_2 & \rightarrow \text{OH + NO} \tag{R6} \label{eq:H_NO2} \\
\text{HO$_2$ + NO} & \rightarrow \text{OH + NO}_2, \tag{R7} \label{eq:HO2_NO} 
\end{align}
%
%O$_2$~+~2H$_2$~$\xrightarrow[]{\text{HO$_x$}}$~2H$_2$O.
%\begin{align}
%\text{H + HO}_2 & \rightarrow \text{OH + OH} \tag{R2} \label{eq:R2}\\
%\text{2(OH + H}_2 & \rightarrow \text{H$_2$O + H)} \tag{R3} \label{eq:R3}\\
%\text{H + O}_2 \text{+ M} & \rightarrow \text{HO$_2$ + M} \tag{R4} \label{eq:R4}\\
%\cline{1-2}
%\text{O$_2$ + 2H$_2$} & \rightarrow \text{2H$_2$O}. \notag
%\end{align}
%The O$_2$ depletion is further enhanced by the presence of large amounts of CO and NO$_x$ (NO$_x$ = NO + NO$_2$):
%\begin{align} 
%\text{H + NO}_2 & \rightarrow \text{OH + NO} \label{eq:R5}  \tag{R5}\\
%\text{2(OH + H}_2 & \rightarrow \text{H$_2$O + H)} \tag{R3}\\
%\text{H + CO + M} & \rightarrow \text{HCO + M} \tag{R6} \label{eq:R6}\\
%\text{HCO + O$_2$} & \rightarrow \text{CO + HO}_2 \tag{R7} \label{eq:R7}\\
%\text{HO$_2$ + NO} & \rightarrow \text{OH + NO}_2 \label{eq:R8} \tag{R8}\\
%\cline{1-2}
%\text{O$_2$ + 2H$_2$} & \rightarrow \text{2H$_2$O}.  \notag
%\end{align}
and by water photolysis (reaction~\ref{eq:H2O_hv}) which increases the concentration of H in the middle atmosphere. The atomic hydrogen can be removed via escape or recombines to form H$_2$ \citep[see][]{hu2012}. 
For the H$_2$-dominated atmosphere with high CH$_4$ concentrations (scenario 1c) the surface mixing ratio of O$_2$ is two orders of magnitudes larger than for scenarios 1a and 1b due to the lower concentrations of NO$_x$ and H in the lower atmosphere. 

For increasing mixing ratios of CO$_2$ the destruction of O$_2$ via H$_2$ oxidation is less dominant and O$_2$ can accumulate in the atmosphere (see Fig.~\ref{fig:evol_prof}).
The CO$_2$-dominated atmospheres with less than 10\% H$_2$ feature large abundances of O$_2$ of up to 30\% for the atmospheres with low and medium CH$_4$ concentrations (scenario 8a--10a and 8b--10b). For the scenarios 8a and 8b the concentration of O$_2$ might be limited to about 10\% (see gray shaded region in Fig~\ref{fig:evol_prof}) due to the potential combustion of the atmosphere \citep{grenfell2018}.

The large concentrations of O$_2$ are related to the lower UV flux of M dwarfs (hence weaker O$_2$ photolysis) as well as strong abiotic production of O$_2$ from CO$_2$ photolysis at wavelengths below 200 nm \citep[see also e.g.][]{domagal2014,harman2015,wunderlich2020}. The lower FUV/NUV ratio of LHS 1140 compared to a solar type star favors the production of abiotic O$_2$ in CO$_2$-rich atmospheres via:
\begin{align} 
\text{2(CO$_2$ + h}\nu & \rightarrow \text{CO + O)} \tag{R8} \label{eq:CO2_hv}\\
\text{O + O + M} & \rightarrow \text{O}_2 \text{+ M} \tag{R9} \label{eq:O_O} \\
\cline{1-2} 
\text{2CO}_2 & \rightarrow \text{2CO + O}_2 \notag 
\end{align}
\citep[see also e.g.][]{selsis2002,segura2007,tian2014,france2016,wunderlich2020}. 
The production of O$_2$ is further enhanced by the presence of OH via:
\begin{align}
\text{O + OH} & \rightarrow \text{O}_2 \text{+ H.}  \tag{R10} \label{eq:O_OH} 
\end{align}
CO and O can recombine to CO$_2$ via a HO$_x$ catalysed reaction sequence, forming CO$_2$: CO~+~O~$\xrightarrow[]{\text{HO$_x$}}$~CO$_2$ \citep[see e.g.][]{domagal2014,gao2015,meadows2017,schwieterman2019}.
%
%The CO$_2$-dominated atmosphere with low CH$_4$ concentrations builds up 3\% of O$_2$ abiotically, assuming a weak O$_2$ deposition of 1$\cdot$10$^{-8}$~cm/s. This is consistent with previous model studies investigating the atmosphere of e.g. the TRAPPIST-1 planets e and f \citep[see]{hu2020,wunderlich2020}. 
%
For the CO$_2$-dominated atmosphere with high CH$_4$ concentrations we find that atomic oxygen (O) quickly reacts with CH$_3$ via:
\begin{align}
    \text{O + CH}_3 & \rightarrow \text{H$_2$CO + H} \tag{R11} \label{eq:O_CH3_1} \\
    & \text{or} \notag \\
    \text{O + CH}_3 & \rightarrow \text{CO + H$_2$ + H,} \tag{R12} \label{eq:O_CH3_2} 
\end{align}
 forming H$_2$, H, CO and H$_2$CO. Part of the H$_2$CO separates into H$_2$ and CO via:
\begin{align}
    \text{H$_2$CO + h$\nu$} & \rightarrow \text{H$_2$ + CO} \tag{R13} \label{eq:H2CO_hv} \\
    & \text{or} \notag \\
    \text{H$_2$CO + H} & \rightarrow \text{H$_2$ + HCO}  \tag{R14} \label{eq:H2CO_H}\\
    \text{HCO + H} & \rightarrow \text{H$_2$ + CO.}  \tag{R15} \label{eq:HCO_H}
\end{align}
Hence, results suggest low concentrations of O$_2$ and large abundances of CO for the high CH$_4$ scenarios with CO$_2$-dominated atmospheres.

\subsubsection{O$_3$}
A main source of O$_3$ largely depends on the amount of O$_2$ available for photolysis in the atmosphere. O$_2$ is split into atomic oxygen via photolysis between 170 and 240~nm (reaction~\ref{eq:O2_hv}) and then reacts with O$_2$ to form O$_3$ via a fast three-body reaction \citep[see e.g.][]{brasseur2006}. The main O$_3$ sinks are photolysis at wavelengths less than $\sim$200~nm and catalytic cycles involving HO$_x$ and NO$_x$ which convert O$_3$ into O$_2$ in the middle atmosphere \citep[see e.g.][]{brasseur2006,grenfell2013}. 

In our scenarios, the low FUV/NUV environment compared with the Earth favors weak release of HO$_x$ and NO$_x$ from their reservoir molecules which leads to weak O$_3$ catalytic loss. 
The low and medium CH$_4$ scenarios with CO$_2$-dominated atmospheres show an ozone layer with peak abundances ($\sim$20--50~ppm) up to about five times larger than on  Earth  \citep[compare to e.g. Fig.~2 of][]{wunderlich2020}.
\citet{grenfell2014} suggested that the smog mechanism is an important O$_3$ source for late M dwarfs since UV levels are not sufficient to drive efficiently the Chapman mechanism.
All scenarios with H$_2$-dominated atmospheres and CO$_2$-dominated atmosphere with high concentrations of CH$_4$ are however low in O$_2$, leading overall to weak production of O$_3$. 

\subsubsection{CO} \label{sec:profiles_co}
All simulations assume emissions of CO from volcanoes and biomass based on pre-industrial Earth values. An important in-situ sink for CO is OH via:
\begin{align}
    \text{CO + OH} & \rightarrow \text{CO$_2$ + H}. \tag{R16} \label{eq:CO_OH}
\end{align}
Our results suggest a decreased OH due to a slowing in its main source reaction:
\begin{align}
    \text{O$^1$D + H$_2$O} & \rightarrow \text{2OH}, \tag{R17} \label{eq:O1D_H2O}
\end{align}
since production of O$^1$D (e.g. from O$_3$ photolysis) is disfavored by the stellar UV emission. The weak OH favors an increase in CO and CH$_4$ by several orders of magnitude compared with modern Earth. Similar effects have been noted by several studies in the literature for Earth-like planets \citep[see e.g.][]{segura2005, grenfell2007, rugheimer2015, wunderlich2019}.

%we obtain increased CO concentrations compared to Earth \citep[see e.g.][]{wunderlich2019}. 
%OH is produced by water photolysis (reaction \label{eq:H2O_hv}) which is much but also by reactions with NO$_x$
The tropospheric temperature of the H$_2$-dominated atmosphere with high concentrations of CH$_4$ (scenario 1c) is much lower compared to the other scenarios due to the strong CH$_4$ anti-greenhouse effect (see above), leading to water condensation hence less water photolysis in the troposphere. 
Note that the OH radical can also be formed via HO$_x$ re-partitioning (reaction~\ref{eq:HO2_NO}), which can be driven by enhancements in NO e.g. via incoming cosmic rays \citep[see e.g.][]{airapetian2016,scheucher2018,airapetian2020}.

%In CO$_2$-dominated atmospheres CO is mainly produced by CO$_2$ photolysis (reaction \ref{eq:R9}). For the low and medium scenario the CO recombines more efficiently to CO$_2$ (via CO~+~O~$\xrightarrow[]{\text{HO$_x$}}$~CO$_2$), resulting in lower CO amounts compared to the other scenario. Therefore, CO is often considered as an "antibiosignature" gas as discussed in e.g. \citet{zahnle2008,wang2016,nava2016,meadows2017,catling2018} and \citet{schwieterman2019}. Note that we use a low deposition velocity of 1$\cdot$10$^{-8}$~cm/s for all simulations, leading to only weak differences of the scenarios \citep[see also][]{schwieterman2019,hu2020,wunderlich2020}. 

In the atmosphere with high concentrations of CH$_4$ results suggest that the recombination of CO and O into CO$_2$ is weakened due to the additional sinks of atomic oxygen via reactions (\ref{eq:O_CH3_1})  and (\ref{eq:O_CH3_2}) as discussed in Section~\ref{sec:profiles_O2}. High CH$_4$ generally favors lowered OH, which weakens the HO$_x$ catalyzed combination of CO and O into CO$_2$. This leads to larger abundances of CO for high CH$_4$ scenarios compared to the other scenarios.

\subsubsection{CH$_4$ and hydrocarbons} \label{sec:profiles_ch4}
Understanding how CH$_4$ forms higher hydrocarbons (C$_n$) and how these species are subsequently removed to reform CH$_4$ is a central issue because higher hydrocarbons can readily condense to form hazes, which could strongly impact climate and observed spectra. 
%As found in previous works the destruction of CH$_4$ is decreased for a planet orbiting an M dwarf compared to a planet around a solar type star due to the reduced sources of OH from e.g. O$_3$ photolysis in the UV \citep[e.g.][]{segura2005, grenfell2013, grenfell2014, rugheimer2015, wunderlich2019}. In H$_2$-dominated atmospheres with low concentrations of O$_3$ the main OH source is  water photolysis (reaction \ref{eq:H2O_hv}).  
%
%. Question: would they condense? What about the long hydrocarbons? Up to  how many carbons  (Cn) does the model simulate
%
The high CH$_4$ scenarios feature a large production of hydrocarbons such as C$_2$H$_2$, C$_2$H$_4$ and C$_2$H$_6$. Figure~\ref{fig:profiles} shows the atmospheric profiles of C$_2$H$_6$ but note that C$_2$H$_2$ and C$_2$H$_4$ (not shown) have similar concentrations with largest mixing ratios at the ToA with up to 0.1\%. 

Our results suggest that in H$_2$-dominated atmospheres the main pathway for initiating ascent of the homologous chain from C$_1$~$\rightarrow$~C$_2$ (CH$_4$ $\rightarrow$ C$_2$H$_6$) pathway is as follows:
\begin{align}
    %\text{CH$_4$ + h}\nu & \rightarrow \text{CH$_3$ + H} \tag{R18}  \label{eq:R18} \\
    \text{2(CH$_4$ + h}\nu & \rightarrow \text{$^1$CH$_2$ + H}_2) \tag{R18} \label{eq:CH4_hv_1} \\
    %\text{2(CH$_4$ + H} & \rightarrow \text{CH$_3$ + H}_2) \tag{R20} \label{eq:R20} \\
    \text{2($^1$CH$_2$ + H}_2 & \rightarrow \text{CH$_3$ + H}) \tag{R19} \label{eq:CH2_H2} \\
    \text{CH$_3$ + CH$_3$ + M} & \rightarrow \text{C$_2$H$_6$ + M} \tag{R20} \label{eq:CH3_CH3} \\
    \cline{1-2}
    \text{2CH$_4$} & \rightarrow \text{C$_2$H$_6$ + 2H}. \notag
\end{align}
The above pathway is an established route for ascending the hydrocarbon chain \citep[see e.g.][Chapter 5 and references therein]{yung1999}. It is initiated by CH$_4$ photolysis to form reactive methyl radicals (CH$_3$), which participate in a three-body self-reaction to form ethane (C$_2$H$_6$). 

Our results suggest that C$_2$H$_4$ is mainly formed by the reactions:
\begin{align}
 \text{CH + CH}_4 & \rightarrow \text{C$_2$H$_4$ + H} \tag{R21} \label{eq:CH_CH4} \\
 & \text{and} \notag \\
 \text{CH$_2$ + CH}_3 & \rightarrow \text{C$_2$H$_4$ + H} \tag{R22}  \label{eq:CH2_CH3} 
\end{align}
where CH is produced via:
\begin{align}
    \text{CH$_4$ + h}\nu & \rightarrow \text{CH + H$_2$ + H} \tag{R23}     \label{eq:CH4_hv_2} \\
    \text{CH$_3$ + h}\nu & \rightarrow \text{CH + H}_2 \tag{R24} \label{eq:CH3_hv} \\
    \text{or} \notag \\
    \text{H + CH}_2 & \rightarrow \text{CH + H}_2 \tag{R25} \label{eq:H_CH2} 
\end{align}
and C$_2$H$_2$ is formed via photolysis of C$_2$H$_4$ and the reaction:
\begin{align}
    \text{CH$_2$ + CH}_2 & \rightarrow \text{C$_2$H$_2$ + H$_2$} \tag{R26} \label{eq:CH2_CH2}
\end{align}
\citep[see also][and references therein]{yung1999}.

In CO$_2$-dominated atmospheres our results suggest that the reaction (\ref{eq:CH2_H2}) is slower due to reduced H$_2$ and since part of the CH$_2$ reacts with CO$_2$ forming H$_2$CO and CO. This effect disfavours the pathway producing C$_2$H$_6$. Additionally however, the destruction of CH$_3$ via:
\begin{align}
    \text{CH$_3$ + H}_2 & \rightarrow \text{CH$_4$ + H} \tag{R27}  \label{eq:CH3_H} 
\end{align}
is weakened due to lowered H$_2$. This effect favors enhanced CH$_3$ hence, the pathway producing C$_2$H$_6$. 
The overall effect is slightly lower concentrations of CH$_4$ in the upper atmosphere
%(slowing in R27, less CH4 production) 
but larger amounts of C$_2$H$_6$ as well as C$_2$H$_2$ and C$_2$H$_4$ (not shown) for CO$_2$-dominated atmospheres compared to H$_2$-dominated atmospheres when assuming high concentrations of CH$_4$.
%(slowing in R27, less CH3 loss which is a source for C2H6)

The larger abundances of C$_4$H$_2$ shown in Figs.~\ref{fig:profiles} and \ref{fig:evol_prof} for the scenario 10c compared to the scenario 1c suggests more haze production by hydrocarbons in CO$_2$-dominated compared with H$_2$-dominated atmospheres. 
Such an effect would be reinforced assuming CO$_2$-dominated atmospheres have cooler mid to upper atmospheres compared with H$_2$-dominated atmospheres as suggested by Fig.~\ref{fig:profiles}. Note however, that the above result could be reversed in atmospheres where CH$_2$ (reaction \ref{eq:CH2_H2}) becomes more important for C$_2$H$_6$ production since H$_2$-dominated atmospheres favor reduced CH$_2$ as discussed above.
For the low and medium CH$_4$ scenarios the concentrations of C$_4$H$_2$ decrease with decreasing CH$_4$/CO$_2$ ratios as suggested by e.g. \citet{arney2018} for N$_2$-dominated atmospheres. 

\subsubsection{SO$_2$} \label{profiles_so2}
We assume volcanic outgassing of SO$_2$ as measured on modern Earth. Due to its high solubility in water forming sulfate, SO$_2$ is deposited easily over wet surfaces leading to SO$_2$ surface mixing ratios below 1~ppb. The main in-situ chemical sink of SO$_2$ is photodissociation below 400 nm \citep[e.g.][]{manatt1993} and its oxidation via reaction with OH or O$_3$ to form ultimately SO$_3$ which quickly reacts with water to form sulfate \citep[see e.g.][]{burkholder2015,seinfeld2016}. In CO$_2$-dominated atmospheres with low and medium CH$_4$ concentrations (scenarios 10a and 10b), results suggest that shielding associated with large UV absorption from CO$_2$, O$_2$, and O$_3$, enables the concentrations of SO$_2$ to reach up to 1~ppm in the stratosphere. 

In H$_2$-dominated atmospheres the high UV environment leads to low abundances of SO$_2$ over the entire atmosphere. Note that strong SO$_2$ abundances in e.g. moist, warm tropospheres (see Figure~\ref{fig:profiles}) would favor significant sulfate aerosol formation although a strong hydrological cycle would quickly wash out the sulfate formed \citep[see e.g.][]{loftus2019}.

\begin{figure*}
\centering
   \includegraphics[width=1.0\textwidth]{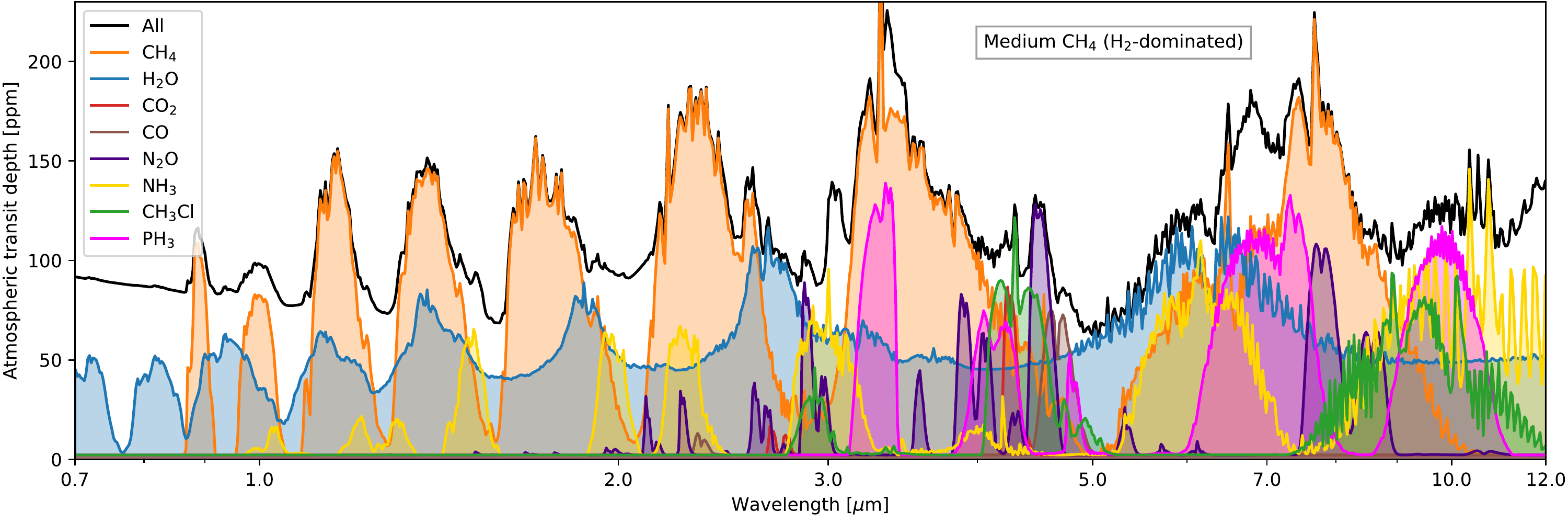}
     \caption{Simulated transmission spectrum of the medium CH$_4$ scenario with the H$_2$-dominated atmosphere (scenario 1b) binned to a constant resolving power of $R$=300 (maximum resolving power of JWST NIRSpec PRISM at 5~$\mu$m). The black line represents the spectrum which includes the contribution from all gases which are considered in 1D-TERRA and also available in HITRAN, Rayleigh scattering, aerosol extinction, H$_2$O CKD and CIAs. Individual gaseous absorption of selected molecules are represented by different colors. }
     \label{fig:tr_contrib}
\end{figure*}

\subsubsection{Potential biosignatures NH$_3$, PH$_3$, CH$_3$Cl and N$_2$O}
%\begin{figure}
%\centering
%   \includegraphics[width=0.9\columnwidth]{04_H2_Profile_comp_biosig}
%     \caption{Composition profiles of NH$_3$, PH$_3$, CH$_3$Cl and N$_2$O for the alive scenario with an H$_2$-dominated atmosophere (solid line) and an CO$_2$-dominated atmosphere (dashed line).}
%     \label{fig:profile_biosig}
%\end{figure}

%In Fig.~\ref{fig:profile_biosig} we show the concentrations of NH$_3$, PH$_3$, CH$_3$Cl and N$_2$O in H$_2$- and CO$_2$-dominated atmospheres with assumed biogenic surface emissions (see Table~\ref{tab:Earth_flux}). The abiotic production of these species is weak, leading to concentrations below 1~ppt. Therefore, we do not further discuss the results of the runs without surface emissions of these species. 

%We consider NH$_3$, PH$_3$, CH$_3$Cl, and N$_2$O as potential biosignature gases in the simulated atmospheres of LHS~1140~b. 
The chemical destruction of NH$_3$, PH$_3$, and CH$_3$Cl is controlled by reactions with OH, O$^1$D, H and by photodissociation in the UV \citep[see e.g.][]{segura2005,seager2013,sousa-silva2020}. 
In the middle atmosphere NH$_3$ photodissociates into NH$_2$ and H. In H$_2$-dominated atmospheres the NH$_2$ reacts quickly with H and reforms NH$_3$. 
In the high CH$_4$ atmosphere this recombination process is slower due to enhanced HO$_2$ from water photolysis in the stratosphere leading to more destruction of atomic hydrogen by reaction (\ref{eq:H_HO2}) (see also Sec.~\ref{sec:profiles_O2}).
Our results suggest that an assumed emission of 8.38$\times$10$^{10}$ molecules~cm$^{-2}$~s$^{-1}$ would lead to NH$_3$ surface mixing ratios between 0.1 and 1~ppm. This is consistent with the results from \citet{seager2013} who obtain an NH$_3$ mixing ratio of 0.1~ppm with surface flux of 5.1$\times$10$^{10}$ molecules~cm$^{-2}$~s$^{-1}$ for a planet with an H$_2$-dominated atmosphere around an active M dwarf.

For PH$_3$ we assume an emissions flux of 1$\times$10$^{10}$ molecules~cm$^{-2}$~s$^{-1}$ and obtain a surface mixing ratio of about 200~ppb for the H$_2$-dominated atmosphere with low CH$_4$ concentrations. \citet{sousa-silva2020} find similar mixing ratios of PH$_3$ when assuming a ten times larger surface flux for a planet with an H$_2$-dominated atmosphere around an active M dwarf. However, they did not consider that PH$_3$ may be recycled via chemical reactions with H or H$_2$. 

For CO$_2$-dominated atmospheres with low and medium concentrations of CH$_4$ (scenarios 10a, 10b) the destruction of PH$_3$ from H is weaker compared to H$_2$-dominated atmospheres and we obtain larger surface mixing ratios of several ppm. Given the weak recycling of PH$_3$ from H or H$_2$, our results are consistent with \citet{sousa-silva2020} who obtain a mixing ratio of 15~ppm for a ten times larger surface flux. For the high CH$_4$ scenario (10c) our results suggest that large abundances of OH are produced near the surface via:
\begin{align}
    \text{CH$_3$ + O}_2 & \rightarrow \text{H$_2$CO + OH}, \tag{R28}     \label{eq:CH3_O2}
\end{align}
leading to enhanced destruction of PH$_3$ compared to the other scenarios. 
The CH$_3$Cl and N$_2$O loss processes are dominated by photolysis in the UV below 240 nm and reaction with O$^1$D \citep[see also e.g.][]{grenfell2013}. Due to the low UV environment in CO$_2$-dominated atmospheres, the destruction by photolysis is weaker and the abundances of CH$_3$Cl and N$_2$O are larger compared to the H$_2$-dominated atmospheres. 

\subsubsection{Atmospheric height} % better scale height?
The right upper panel of Fig.~\ref{fig:evol_prof} shows atmospheric height at the ToA for all simulated scenarios. A central aim of this paper is to investigate whether it is feasible to detect atmospheric molecular features on LHS~1140~b with future
telescopes. A large atmospheric height leads to stronger absorption features in transmission spectroscopy and hence to improved
detectability of the corresponding species. Due to the low mean molecular weight hence larger scale height of the H$_2$-dominated atmospheres, the ToA
at $\sim$0.01 Pa occurs at a height of up to 570~km, whereas the CO$_2$-dominated atmospheres only reach an altitude of about 35~km. The larger
stratospheric temperatures for the high CH$_4$ scenarios furthermore lead to an expansion of the atmosphere compared to the scenarios with less CH$_4$.

\subsection{Transmission spectra}

\begin{table}
\centering
\caption{Central wavelength, $\lambda_{\text{c}}$ ($\mu$m), of molecular bands from NH$_3$, PH$_3$, CH$_3$Cl, and N$_2$O and spectral features which might overlap or obscure the bands at the corresponding $\lambda_{\text{c}}$. }
\label{tab:features}     
\centering                                     
\begin{tabular}{lcc}        
\hline\hline 
    Species &  $\lambda_{\text{c}}$ &  Overlap  \\
    \hline 
N$_2$O   &  2.9~$\mu$m  &  CO$_2$ \\
N$_2$O   &  4.5~$\mu$m  &   CO, CO$_2$ \\
NH$_3$   &  2.0~$\mu$m  &   H$_2$O, H$_2$-H$_2$, haze, CO$_2$ \\
NH$_3$   &  3.0~$\mu$m  &   CO$_2$, C$_2$H$_2$ \\
NH$_3$   &  6.1~$\mu$m   &  CH$_4$, H$_2$O \\
NH$_3$   &  10.5~$\mu$m  &  H$_2$-H$_2$, O$_3$, CO$_2$ \\
PH$_3$   &  4.3~$\mu$m   &  CO$_2$ \\
PH$_3$   &  9.5~$\mu$m   &  H$_2$-H$_2$, O$_3$, CO$_2$ \\
CH$_3$Cl &  3.3~$\mu$m   &  CH$_4$ \\
CH$_3$Cl &  7.0~$\mu$m   &  CH$_4$, H$_2$O, C$_2$H$_6$ \\
CH$_3$Cl &  9.8~$\mu$m   &  H$_2$-H$_2$, O$_3$, CO$_2$ \\
    \hline                                            
\end{tabular}
\end{table}

\begin{figure*}
\centering
   \includegraphics[width=1.0\textwidth]{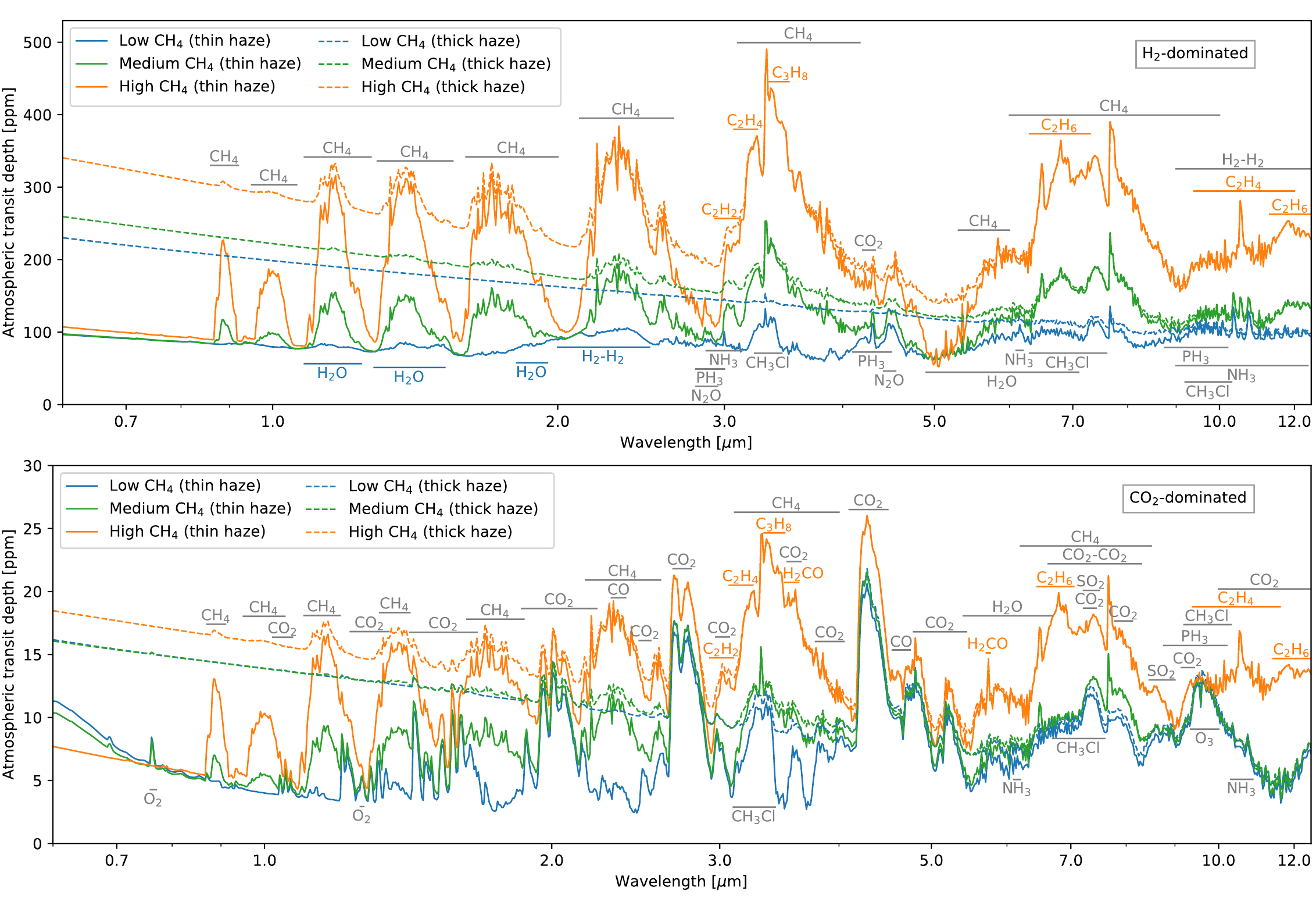}
     \caption{Simulated atmospheric features of the LHS~1140~b for all three scenarios with H$_2$- (upper panel) and CO$_2$-dominated atmospheres (lower panel), represented by cloud-free transit transmission spectra and binned to a constant resolving power of $R$=300. The scenarios are indicated by the different colors: blue for low concentrations of CH$_4$ (scenario 1a for H$_2$- and scenario 10a for CO$_2$-dominated atmospheres), green for medium concentrations of CH$_4$ (scenario 1b for H$_2$- and scenario 10b for CO$_2$-dominated atmospheres) and orange for high concentrations of CH$_4$ (scenario 1c for H$_2$- and scenario 10c for CO$_2$-dominated atmospheres). Important atmospheric molecular absorption are highlighted with horizontal lines in the color of the scenario with a strong absorption feature or in gray when all scenarios show a strong feature. Solid lines include weak extinction from hazes whereas dashed lines consider extinction by thick hazes.}
     \label{fig:tr_spec}
\end{figure*}

Figure~\ref{fig:tr_contrib} shows the simulated transmission spectrum of LHS~1140~b for the H$_2$-dominated atmosphere with medium CH$_4$ concentrations (scenario 1b). The contribution from individual molecular absorption bands is shown with different colors. To detect a spectral band with e.g. the JWST it is important to identify a wavelength range which is not obscured by the absorption of other molecules or haze extinction. 
The strongest spectral features are due to absorption by CH$_4$. Below $\sim$2~$\mu$m molecular features of H$_2$O and NH$_3$ are obscured by haze extinction, which increases the transit depth to $\sim$90~ppm. Between 2 and 2.5~$\mu$m H$_2$-H$_2$ CIA contributes significantly to the transmission spectrum \citep[see][]{abel2011}.

Around 3.0~$\mu$m we obtain two absorption features mainly produced by NH$_3$, N$_2$O and PH$_3$. Absorption by CH$_4$ and H$_2$O is rather weak at these wavelengths. However, between 3.0 and 3.1~$\mu$m C$_2$H$_2$ contributes significantly to the spectral feature (not shown). The absorption by CO$_2$ around 3.0~$\mu$m is significant for atmospheres with mixing ratios of $\sim$1\% or more CO$_2$ (not shown). 
The spectral features of CH$_3$Cl around 3.3~$\mu$m and 7~$\mu$m overlap with absorption by CH$_4$ and H$_2$O. 
For CO$_2$-poor atmospheres the features from N$_2$O and PH$_3$ dominate the spectrum between 4.1 and 4.6~$\mu$m. However, Earth-like CO$_2$ levels lead to a strong absorption feature around 4.3~$\mu$m  \citep[see e.g.][]{rauer2011,wunderlich2019}. 

Between 9 and 12~$\mu$m we find strong absorption by PH$_3$, NH$_3$ and CH$_3$Cl. For H$_2$-dominated atmospheres there is a significant contribution from H$_2$-H$_2$ CIA (not shown) to the transmission spectrum \citep{abel2011,fletcher2018}. In CO$_2$ atmospheres these features might overlap with absorption by O$_3$ and CO$_2$. 
Many of the simulated atmospheric spectral features of N$_2$O, NH$_3$, PH$_3$ or CH$_3$Cl overlap with other molecular bands of e.g. CO$_2$ or CH$_4$ (see Table~\ref{tab:features}). However, at some spectral bands these potential biosignatures contribute significantly to the full feature offering the possibility to detect the additional absorption if the abundances of CO$_2$ and CH$_4$ are known.
%In Section~\ref{sec:snr} we calculate the number of transits necessary for selected cases 
Figure~\ref{fig:tr_spec} shows the simulated transmission spectra of H$_2$-dominated atmospheres (scenarios 1a, 1b, and 1c) and CO$_2$-dominated  atmospheres (scenarios 10a, 10b, and 10c). 
%The spectra are simulated by the GARLIC model taking as input the chemical and temperature profiles discussed in Section~\ref{sec:profiles}. 
We take into account the effect of weak extinction from thin hazes (solid line) and from thick Titan-like hazes (dashed line).  

%In H$_2$-dominated atmosphere the CH$_4$ features are dominant for all scenarios with different intensities. For the low CH$_4$ atmosphere the H$_2$-H$_2$ CIA is visible between 2.0 - 2.5~$\mu$m. For the other two scenarios strong absorption by CH$_4$ overlaps with the  H$_2$-H$_2$ CIA band. CO$_2$ shows only a weak absorption feature at 4.3 $\mu$m. The high CH$_4$ has the largest spectral features reaching almost 500~ppm around 3.2 $\mu$m.  

For the high CH$_4$ scenarios the mean transit depth is increased compared to the other two scenarios due to strong absorption by CH$_4$ and the warm stratospheric temperature which leads to an expansion of the atmosphere.
Due to the low molecular weight of the H$_2$-dominated atmospheres the spectral features are generally larger than in CO$_2$-dominated atmospheres. 
The extinction by thick haze significantly increases the transit depth at atmospheric windows up to 6~$\mu$m. 
Hazes at large altitudes are considered to be the main reason for the observed flat spectrum of e.g. GJ~1214~b \citep{bean2010, desert2011, kempton2011, kreidberg2014}. However, HST observations of the atmosphere of LHS~1140~b suggest a clear atmosphere \citep{edwards2020}. In the following text we compare the observed spectrum with our simulations.

% begin at 0.7
% NH3 is not visible for H2-dom at 6mum
% CO2 only for low CH4
% overlap of nh3, ph3, n2o enhace feature, difficult to distinguish with lrs but hints at biosignature

%- PH3: for CO2 rich atmosphere features around 2.9 and 4.4 are not visible due to CO2 absorption
%- C2h4 and C2H6 overlap with CH4 bands

\subsection{Comparison to observations}
\begin{figure*}
\centering
   \includegraphics[width=1.0\textwidth]{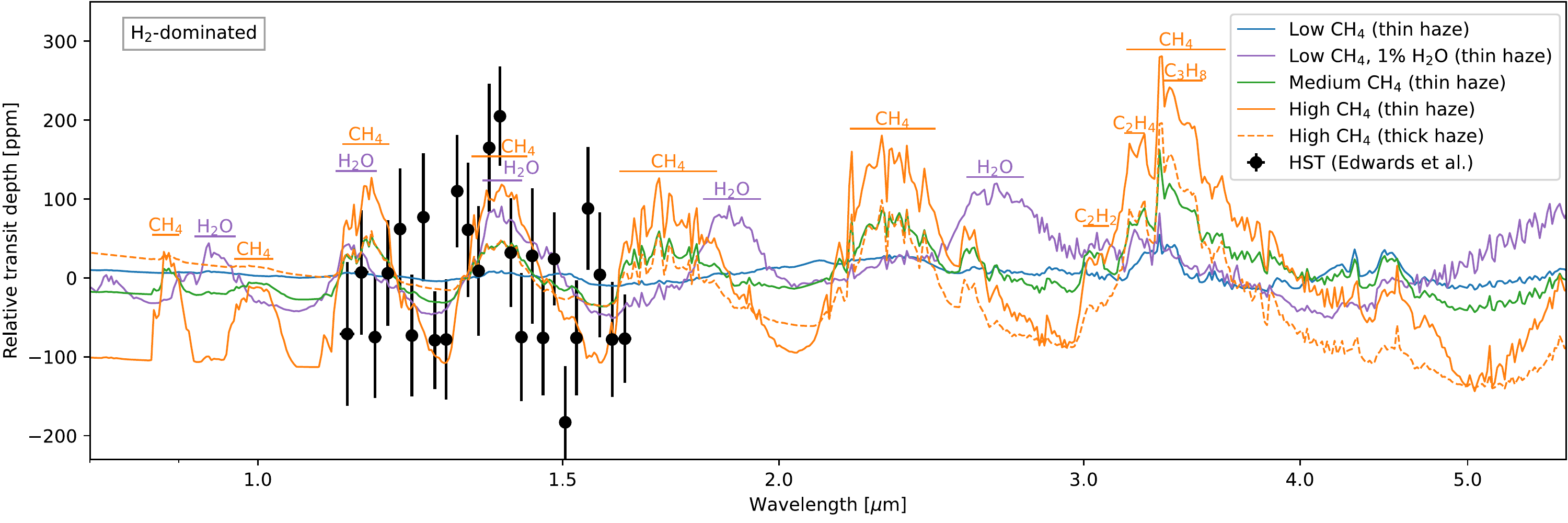}
     \caption{Simulated spectral features compared to HST observations taken from \citet{edwards2020}, assuming H$_2$-dominated atmospheres. 
     The scenarios are indicated by the different colors: solid blue line for the low CH$_4$ scenario (1a) as shown in Fig.~\ref{fig:tr_spec} and the solid purple line for scenario 1a but with an assumed constant H$_2$O mixing ratio of 1\%; solid green line for the medium CH$_4$ scenario (1b); solid and dashed orange lines for the high CH$_4$ scenario (1c) with thin and thick hazes, respectively. All spectral data were binned to $R$=300 and subtracted from the mean transit depth between 1.1 and 1.7~$\mu$m.}
     \label{fig:tr_obs}
\end{figure*}

\citet{diamond2020} combined two spectrally resolved transit observations of LHS~1140~b from the optical to the NIR (610--1010~nm). Their median uncertainty of the transit depth was 260~ppm. They concluded that about a factor of 4 higher precision would be needed to detect a clear hydrogen-dominated atmosphere. 
The strongest spectral feature of CH$_4$ at $\sim$900~nm has a strength of about 130~ppm assuming weak extinction by hazes (solid orange line in Fig.~\ref{fig:tr_spec}). The slope from extinction by thick hazes between 610 and 1010~nm leads to a decrease of the transit depth by $\sim$50~ppm (dashed orange line in Fig.~\ref{fig:tr_spec}, upper panel). Hence, our results confirm that the precision of transit observations between 610 and 1010~nm shown by \citet{diamond2020} is not large enough to draw a conclusion on the atmosphere of LHS~1140~b. 

Recently, \citet{edwards2020} presented HST transit observations between 1.1 and 1.7~$\mu$m. 
They concluded that a maximum in the spectrum around 1.4~$\mu$m might suggest an evidence of water vapour absorption in a clear H$_2$/He atmosphere. However, due to the large stellar contamination and the low overall signal-to-noise ratio further observation time is required to confirm the detection of a planetary atmospheric feature. 
Figure~\ref{fig:tr_obs} compares the spectrally resolved HST observations with simulated spectra assuming H$_2$-dominated atmospheres (scenarios 1a, 1b, and 1c). The low CH$_4$ scenario shows a mixing ratio of H$_2$O below 1~ppm  in the middle atmosphere (see Fig.~\ref{fig:profiles}) and the resulting spectrum shows only a weak feature at 1.4~$\mu$m (blue line). In \citet{edwards2020} the water vapour abundance has been retrieved to log$_{10}$(V$_{\text{H$_2$O}}$)~=~-2.94$^{+1.45}_{-1.49}$. 
When assuming a H$_2$O mixing ratio of 1\%, constant over height, we find a difference in the transit depth of about 150~ppm between 1.4~$\mu$m and 1.6~$\mu$m. However, such large abundances of H$_2$O are not consistent with the results of our photochemical model simulations for thin H$_2$-dominated atmospheres with habitable surface temperatures. Large abundances of H$_2$O in thick H$_2$-atmospheres, which were not considered here, might be consistent with our model but would lead to surface temperatures above 395~K (see Fig.~\ref{fig:tsurf}), which are unlikely to sustain life \citep[see e.g.][]{bains2015}. 

Figure~\ref{fig:tr_obs} suggests that a large spectral feature of $\sim$200~ppm at 1.4~$\mu$m can be also obtained by strong CH$_4$ absorption in a thin H$_2$-dominated atmosphere containing several percent of CH$_4$ and limited haze production. Large abundances of CH$_4$ might favor the formation of hydrocarbon haze \citep{he2018,horst2018,lavvas2019}. However, Fig.~\ref{fig:profiles} suggests that the haze production is lower compared to CO$_2$-dominated atmospheres with high CH$_4$ abundances (see Section~\ref{sec:profiles_ch4}). 

Similarly to our finding regarding the atmosphere of LHS~1140~b, the model studies by \citet{bezard2020} and \citet{blain2020} suggest that the detected spectral feature at 1.4~$\mu$m in the atmosphere of K2-18~b \citep{benneke2019,tsiaras2019} might be produced by CH$_4$ rather than H$_2$O. They concluded that the H$_2$O-dominated spectrum interpretation is either due to the omission of CH$_4$ absorption or a strong overfitting of the data.
Further observations with e.g. the Very Large Telescope (VLT) or the JWST are expected to confirm or rule-out the existence of large abundances of CH$_4$ in the atmosphere of LHS~1140~b or K2-18b \citep[see e.g.][]{edwards2020,blain2020}.
%
%The V-band magnitude of LHS~1140 is 14.150 \citep{zacharias2013} and hence, much weaker than the K-band magnitude of 8.821 \citep{cutri2003}. \citet{diamond2020} show that it is challenging to detect an atmosphere on LHS~1140~b with state of the art telescopes. 
In the following we further determine the capabilities of the upcoming generation of telescopes with increased sensitivity and larger wavelength coverage to characterize the atmosphere of LHS~1140~b.

%New fig: as Fig. 6 w/o obs and w/o 1%H2O
% spectrum full and spectrum - N2O, CH3Cl, PH3, NH3

\subsection{Detectability of spectral features with JWST and ELT}
\label{sec:detection}

\begin{figure*}
\centering
   \includegraphics[width=1.0\textwidth]{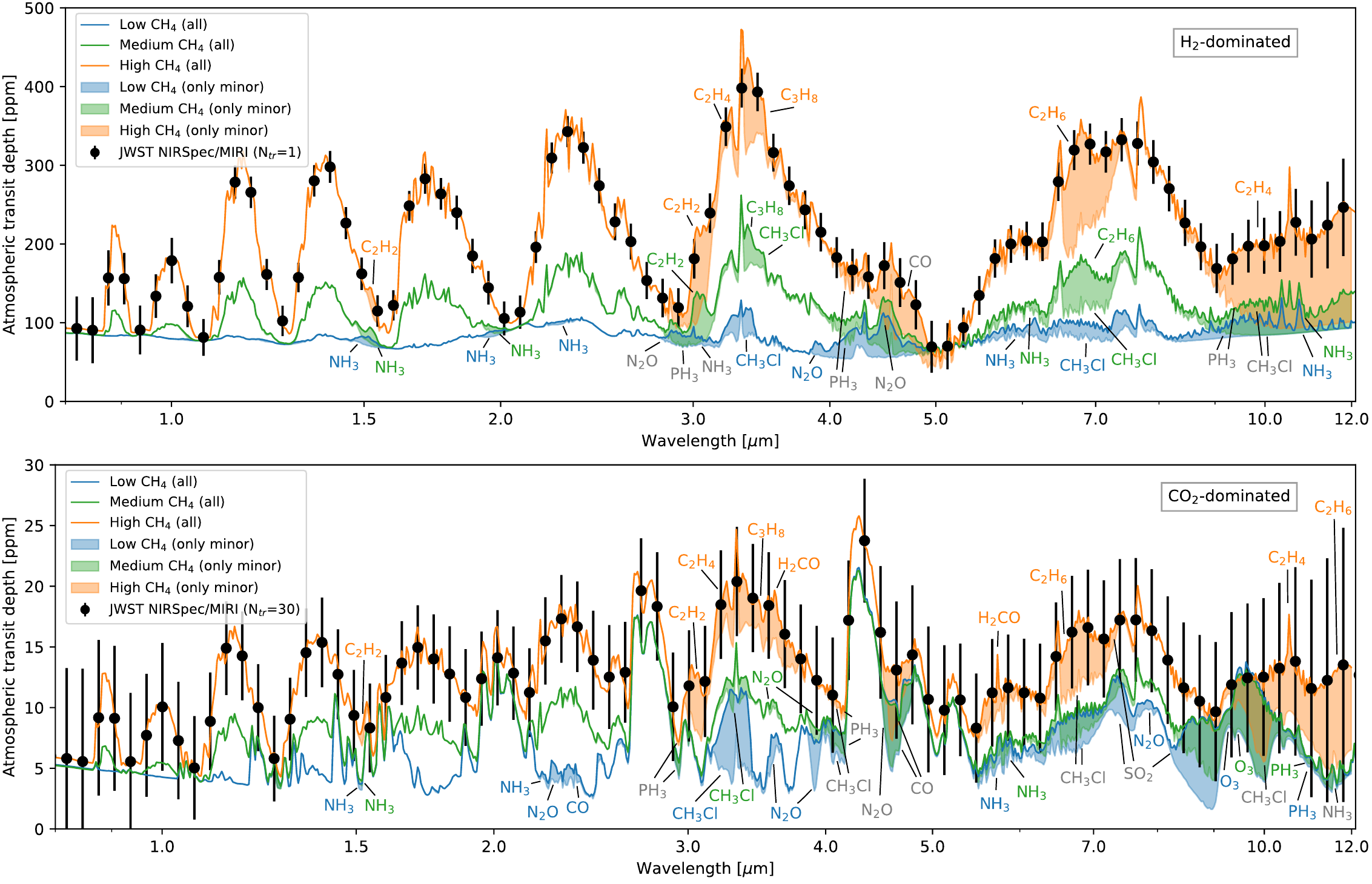}
     \caption{Simulated transmission spectra of LHS~1140~b with weak extinction from hazes in solid lines as shown in Fig.~\ref{fig:tr_spec}. The upper panel shows H$_2$-dominated atmospheres with low, medium and high CH$_4$ concentrations corresponding to scenarios 1a (blue), 1b (green), and 1c (orange), respectively. The lower panel shows CO$_2$-dominated atmospheres with low, medium and high CH$_4$ concentrations corresponding to scenarios 10a (blue), 10b (green), and 10c (orange), respectively. The shaded region shows the contribution of the minor species to the full spectrum.
     Expected error bars for a single transit observation assuming scenario 1c (upper panel) and 30 co-added transits assuming scenario 10c (lower panel) using JWST NIRSpec PRISM (0.7--5~$\mu$m) and JWST MIRI LRS (5--12~$\mu$m), binned to $R$=30. Strongest contribution of minor atmospheric molecular absorption bands are indicated by the color of the scenario or in gray when all scenarios have a significant contribution.}
     \label{fig:tr_minor}
\end{figure*}

\begin{figure*}
\centering
   \includegraphics[width=1.0\textwidth]{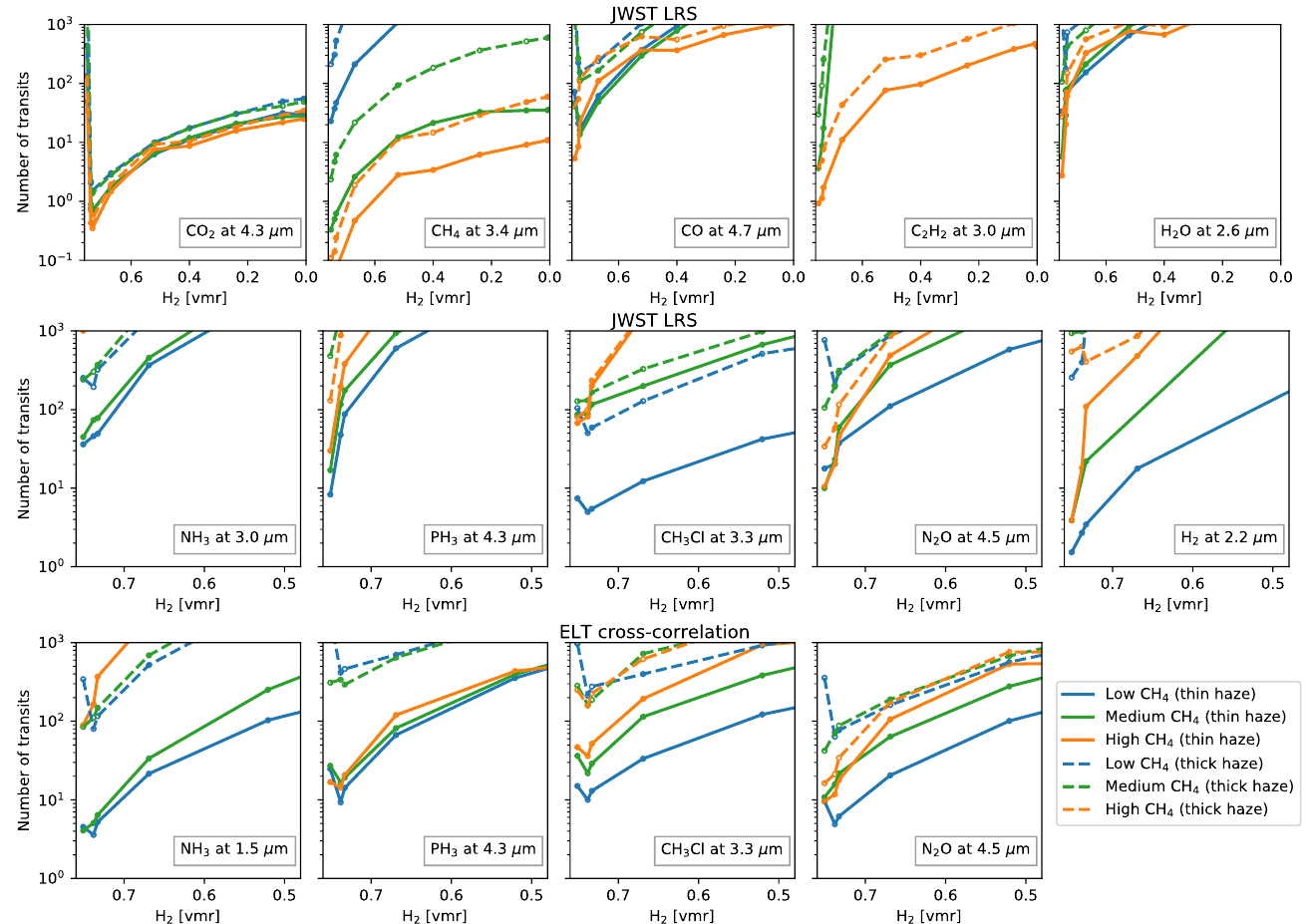}
     \caption{Number of transits required to reach an S/N of 5 for the corresponding spectral features in a cloud free atmosphere with thin (solid lines) and thick (dashed lines) hazes using low resolution transmission spectroscopy with JWST NIRSpec PRISM (upper and middle panels) and the cross-correlation technique with high resolution transmission spectroscopy using ELT HIRES or METIS (lower panels). Note that the x-axis of the upper panel is different to the middle and lower panel. Each colored circle corresponds to one scenario from Table~\ref{tab:composition}. Blue lines and circles show low CH$_4$ scenarios, green lines and circles show medium CH$_4$ scenarios and orange lines and circles show high CH$_4$ scenarios.}
     \label{fig:ntr}
\end{figure*}

%\begin{table}
%\centering
%\caption{Central wavelength, $\lambda_{\text{c}}$ ($\mu$m), of molecular bands from N$_2$O, NH$_3$, PH$_3$ and CH$_3$Cl and spectral features which might overlap or obscure the bands at the corresponding $\lambda_{\text{c}}$. }
%\label{tab:ntr}
%\centering                                     
%\begin{tabular}{lcccc}        
%\hline\hline 
%    Species &  Instr. &  Low CH$_4$ &  Medium CH$_4$ & High CH$_4$\\
%    \hline 
%NH$_3$   &  JWST  &  35 (-) & 37 (-) & - (-) \\
%NH$_3$   &  ELT  &  5 (232) & 4 (96) & 90 (-) \\
%PH$_3$   &  JWST  &  9 (-) & 12 (689) & 30 (131) \\
%PH$_3$   &  ELT  &  27 (-) & 27 (-) & 17 (-) \\
%CH$_3$Cl   &  JWST  &  29 (-) & 140 (472) & 409 (474) \\
%CH$_3$Cl   &  ELT  &  36 (938) & 43 (206) & 47 (250) \\
%N$_2$O   &  JWST  &  18 (671) & 11 (159) & 11 (35) \\
%N$_2$O   &  ELT  &  13 (310) & 11 (63) & 10 (17) \\
%    \hline                                            
%\end{tabular}
%\end{table}

To determine the number of transits which are required to detect a spectral feature (S/N = 5) we subtract the full transmission spectra (including the absorption from all species, CIAs, H$_2$O CKD, Rayleigh extinction and extinction from hazes) from the spectrum excluding the contribution from the corresponding species. Thus, we consider only the contribution of these species to the full spectrum. This method assumes that we know the concentration of other main absorber such as CH$_4$ or CO$_2$, which may overlap with the spectral bands (see Table~\ref{tab:features}). Note that other molecules could mimic large scale features of the molecule in question. Hence, retrieval analysis \citep[see][for a recent overview]{barstow2020} or the detection of the molecules at multiple wavelengths would be required to exclude an ambiguity.  

Second, for LRS with JWST NIRSpec we bin the spectral data until the optimal value is found, leading to the lowest required number of transits. Binning the data decreases on the one hand the noise contamination but on the other hand, large wavelength ranges can lead to interfering overlaps of absorption bands and atmospheric windows. 
Due to the unknown systematic error when binning the synthetic spectral data we assume only white noise for the binning. This gives an optimistic estimation on the detection feasibility of the JWST.
Note that we do not consider the wavelength range between 1 and 2~$\mu$m due to the saturation of the detector for NIRSpec PRISM. However, due to extinction by hazes the detection of the spectral features of e.g. CH$_4$ and CO$_2$ at this wavelength range would require similar or more observation time than the features at longer wavelengths.
For the high resolution spectra we use a constant resolving power of $R$=100,000 as planned for the HIRES and METIS on ELT and apply the cross-correlation technique with the same method as presented in \citet{wunderlich2020}.  
We do not show results from emission spectroscopy in this study due to the much larger required observation time compared to transmission spectroscopy for planets in the habitable zone \citep[see e.g.][]{rauer2011,lustig-yaeger2019}. 

At most wavelengths the simulated spectra are dominated by major absorbing species (here defined as H$_2$, He, CO$_2$, CH$_4$, and H$_2$O). To identify suitable wavelength ranges for the detection of minor absorbers (here defined as all non-major species), the spectrum including only major absorbers was subtracted from the full spectrum. The remainder, representing the contribution of the minor absorbers to the full spectrum, is shown in the shaded region in Fig.~\ref{fig:tr_minor}. The expected error bars of the simulated H$_2$-dominated atmospheres observed by JWST NIRSpec PRISM or MIRI LRS suggest that a detection of minor absorbers will be challenging within a single transit (upper panel of Fig.~\ref{fig:tr_minor}). However, multiple transits might improve the S/N sufficiently to detect those features. 
%Regarding potential biosignatures, in particular the wavelength range between 2.8 and 4.8~$\mu$m and large parts of the spectrum covered by MIRI LRS are favored. For the high CH$_4$ scenario there wavelength ranges would overlap with absorption by hydrocarbons. 

In CO$_2$-dominated atmospheres the larger molecular weight decreases the features in the transmission spectra compared to H$_2$-dominated atmospheres.
The error bars in the lower panel of Fig.~\ref{fig:tr_minor} arise from 30 co-added transit observation with JWST, which would correspond to a period of two years if each transit of LHS~1140~b were to be observed. The results suggest that it will be very challenging to detect minor absorbers in a CO$_2$-dominated atmosphere during the lifetime of JWST. The detection of major absorbers such as CO$_2$ or CH$_4$ might be feasible but also challenging. This is consistent with the results of \citet{morley2017}, who suggest that the detection of a Venus-like atmosphere on LHS~1140~b would require over 60~transits with JWST.

The upper panel of Fig.~\ref{fig:ntr} suggests that CO$_2$ is detectable at 4.3~$\mu$m for mixing ratios between $1\times10^{-3}$ and 0.1 within a few transits. For the scenarios 1a, 1b, and 1c with only 1~ppb CO$_2$ the spectral features are too weak to allow for a detection of CO$_2$ (see also Fig.~\ref{fig:tr_contrib}).  
The detection of CO$_2$ is only weakly dependent on the amount of CH$_4$ in the atmosphere. The high CH$_4$ cases require less transits due to the larger atmospheric heights from CH$_4$ heating in the middle atmosphere (see Fig.~\ref{fig:evol_prof}). The detection of CH$_4$ in a clear H$_2$-dominated atmosphere requires only one transit observation with CO$_2$ concentrations of less than 10\% for the high CH$_4$ scenarios and with less than 1\% CO$_2$ for the medium CH$_4$ scenarios. When assuming thick hazes the detection of CH$_4$ would require about ten times more transits for the medium and low CH$_4$ scenarios. For the high CH$_4$ scenarios the impact of haze extinction on the detectability of CH$_4$ is weaker. 
For CO$_2$-dominated atmosphere the detection of the 3.4~$\mu$m CH$_4$ feature will be challenging with transmission spectroscopy. 
The detectability of CH$_4$ and CO$_2$ is not significantly improved when applying the cross-correlation technique with observations by ELT HIRES or METIS \citep[not shown, see e.g.][]{wunderlich2020}.  

For each of the atmospheric scenarios we assume a strong dry deposition of CO (see Table~\ref{tab:Earth_flux}) leading to weak accumulation of CO from CO$_2$ photolysis in CO$_2$-rich atmospheres compared to previous model studies \citep{schwieterman2019,hu2020,wunderlich2020}. Hence, the detection of CO will be challenging in the atmospheres we consider. 
Hydrocarbons such as C$_2$H$_2$ are formed in large abundances for the high CH$_4$ case and could be detectable with less than 10 transits in CO$_2$ poor, H$_2$-dominated atmospheres. 

The number of transits required to detect H$_2$O is lowest for the H$_2$-dominated atmosphere with high CH$_4$ concentration (scenario 1c). This scenario has the lowest surface water vapor but large amounts of chemically produced H$_2$O in the stratosphere. Hence, the detection of water in transmission spectroscopy is only weakly related to the presence of liquid surface water in these cases.

In the middle panel of Fig.~\ref{fig:ntr} we show the number of transits required to detect spectral features from NH$_3$, PH$_3$, CH$_3$Cl, N$_2$O, and the CIA from H$_2$-H$_2$ using JWST NIRSpec PRISM. The lower panel of Fig.~\ref{fig:ntr} shows the number of transits required to detect spectral features from NH$_3$, PH$_3$, CH$_3$Cl, and N$_2$O using ELT HIRES. Note that we show only the scenarios with H$_2$ mixing ratios of more than 50\%. For CO$_2$-poor atmospheres with low or medium concentrations of CH$_4$ (scenarios 1a--3a and 1b--3b) the results suggest that the detection of NH$_3$ would require tens to hundreds of transits with JWST NIRSpec PRISM and the spectral lines of NH$_3$ around 1.5~$\mu$m might be detectable with about five transits with ELT HIRES. The detection of NH$_3$ around 2.3~$\mu$m would require around 8 transits (not shown). However, due to possibility of a simultaneous detection of spectral lines from CH$_4$, H$_2$O and CO this wavelength region might be favorable for HRS \citep{brogi2019}. Strong extinction by hazes or large absorption by CH$_4$ for the high CH$_4$ scenarios would prevent the detection of NH$_3$.

The spectral feature of PH$_3$ at 4.3~$\mu$m might be detectable within 10--30 transits for the H$_2$-dominated atmospheres with 1~ppb CO$_2$ (scenarios 1a, 1b, and 1c) and weak extinction by hazes. For the other scenarios the PH$_3$ feature is obscured by the CO$_2$ absorption band around 4.3~$\mu$m  \citep[see also][]{sousa-silva2020}. 
The detectability of PH$_3$ is improved using high resolution spectra of ELT METIS compared to JWST observations for CO$_2$-rich atmospheres. However, for more than 1\% of CO$_2$ mixing ratios the detection of PH$_3$ would require tens to hundreds of transits.

For H$_2$-dominated atmospheres with high concentrations of CH$_4$ (scenario 1c) the mixing ratios of CH$_3$Cl are larger compared to scenarios 1a and 1b (see Fig.~\ref{fig:profiles}). However, the strongest spectral band of CH$_3$Cl in the wavelength range of JWST NIRSpec PRISM overlaps with absorption by CH$_4$ (see Fig.~\ref{fig:tr_contrib}), owing to a better detectablility of CH$_3$Cl in CH$_4$-poor atmospheres. 
Similar observation time is required to detect CH$_3$Cl with JWST and with ELT for the low CH$_4$ scenarios. However, the cross-correlation technique is less sensitive to the increase in CH$_4$ compared to LRS. Note that 10 transit observations with JWST would be feasible within one or two years \citep[given a 24.7 days orbital period of LHS~1140~b,][]{dittmann2017} but ground-based facilities would require a much longer observation period because less transits could be captured per year. 
N$_2$O might be detectable within 10 to 20 transits in CO$_2$-poor atmospheres with thin hazes. Results suggest weak dependence of the detectability of N$_2$O on the concentration of CH$_4$. 

%Our paper suggests that the detection of the potential biosignatures investigated will be challenging with JWST or ELT even in an observationally advantageous H$_2$-dominated atmosphere.  
%Hence, retrievals will also challenge to constrain the abundance with low uncertainty due to the low S/N and sparse knowledge of broadening coefficients.
%For the high CH$_4$ scenarios N$_2$O might be detectable for CO$_2$-poor atmospheres with thick hazes.
Our paper suggests that the detection of potential biosignatures with JWST or ELT is feasible for clear, H$_2$-dominated atmosphere but would require several transit observations. If such a molecule would be detected, retrieval analysis might challenge to constrain its abundance with low uncertainties due to sparse knowledge on broadening coefficients \citep[see e.g.][]{Tennyson15,Hedges16,Barton17b,Fortney19} and hence, it would be difficult to rule out a potential abiotic origin. 

\section{Summary and Conclusion} \label{sec:summary}
In this study we used a self-consistent model suite to simulate the atmosphere and spectral appearance of LHS~1140~b.
First we performed climate only runs to determine the surface pressures for which the Super-Earth LHS~1140~b would have habitable surface conditions in N$_2$, H$_2$ and CO$_2$ atmospheres. Our results suggest that a thick N$_2$-dominated atmosphere on LHS-1140~b or substantial amounts of greenhouse gases such as CO$_2$ would be required to reach habitable surface temperatures. A $\sim$2.5~bar CO$_2$ atmosphere or a $\sim$0.6~bar H$_2$-He atmosphere would lead a surface temperature of $\sim$273~K. 
In a second step we used these results and assumed a fixed surface pressure of 2.416~bar (corresponding to the atmospheric mass of the Earth) to simulate potential CO$_2$- and H$_2$-dominated atmospheres of LHS~1140~b with our coupled climate-photochemistry model, 1D-TERRA. We simulated possible composition of the planetary atmospheres, assuming fixed biomass emissions and varying boundary conditions for CH$_4$. 

The results suggest that the amount of atmospheric CH$_4$ can have a large impact upon the temperature and composition of H$_2$-dominated atmospheres. 
A few percent of CH$_4$ may be enough to lower the surface temperatures due to an anti-greenhouse effect. In H$_2$-dominated atmospheres with high concentrations of CH$_4$ this effect dominates, leading to a cooling of up to 100~K and the stratosphere is pronounced with temperatures up to 70 K warmer than those at the surface.
Although we did not consider the effect of hydrocarbon hazes in the climate-chemistry model, e.g. \citet{arney2017} have shown that this is expected to warm the surface temperature by only a few degrees.
For CO$_2$ atmospheres the temperature profile is less affected by CH$_4$ absorption due to CO$_2$ cooling in the stratosphere.

In H$_2$-dominated atmospheres O$_2$ is efficiently destroyed preventing significant concentrations of O$_2$ and O$_3$ in such environments. Hence, even if O$_2$ and O$_3$ were biosignatures, they would not be detectable in the atmosphere of such a habitable planet which is dominated by H$_2$. In CO$_2$-dominated atmospheres O$_2$ and O$_3$ can be produced abiotically which might lead to a false-positive detection \citep[see also][]{selsis2002, segura2007, harman2015, meadows2017, wunderlich2020}.  However, results suggest that large amounts of CH$_4$ would also lead to low concentrations of O$_2$ and O$_3$.  

We consider NH$_3$, PH$_3$, CH$_3$Cl and N$_2$O to be potential biosignatures in H$_2$ and CO$_2$ atmospheres. Here the main constituent of the atmosphere has a weak impact on the concentrations of these potential biosignatures assuming that the emission flux is the same for both H$_2$ and CO$_2$ atmospheres \citep[see also][]{seager2013,sousa-silva2020}. However, the detectability of molecules with transmission spectroscopy largely depends on the main composition of the atmosphere due to the difference in mean molecular weight. 
%Our results suggest that if LHS~1140~b does not have an atmosphere dominated by H$_2$ or He, at least 100 hours of JWST observation time are required to detect spectral features of CO$_2$ or CH$_4$ with transmission spectroscopy \citep[see also][]{morley2017}. 

First observations of the planet suggest that the atmosphere of LHS~1140~b has a low mean molecular weight \citep{edwards2020}. In a thin, H$_2$-dominated atmosphere our results suggest that the tentative spectral feature at 1.4~$\mu$m might be produced by CH$_4$ rather than H$_2$O. If the feature at 1.4~$\mu$m were produced by water vapour absorption, surface temperatures are unlikely to be habitable. Our results suggest that the molecular features of CH$_4$ and CO$_2$ for habitable surface conditions might be detectable within one transit using JWST NIRSpec observations around 3.4~$\mu$m and 4.3~$\mu$m, respectively. At these wavelengths the absorption cross section of H$_2$O is weak and at large wavelengths the extinction by hazes has only a weak impact on the detectability of spectral features.
The detection of NH$_3$, PH$_3$, CH$_3$Cl or N$_2$O would require about 10--50 transits ($\sim$40--200~h) with JWST, assuming clear conditions. 
The molecular bands of these species overlap with absorption by CO$_2$ or CH$_4$ in most cases, making a detection more challenging.  

With high resolution spectroscopy using ELT HIRES or METIS individual absorption lines are distinguishable which might improve the detectability of potential biosignatures. Our results suggest that NH$_3$ might be detectable with less than 20~h of ELT observing time in H$_2$-dominated atmospheres with low or medium CH$_4$ mixing ratios.  
A thick haze layer in the atmosphere would, however prevent the detection of any potential biosignature. Strong spectral lines of PH$_3$, CH$_3$Cl and N$_2$O feature in the wavelength range of ELT METIS with overall lower sensitivity compared to ELT HIRES \citep[see e.g.][]{wunderlich2020}.

Results suggest that a single transit observation of LHS~1140~b with JWST NIRSpec would be enough to confirm or rule out the existence of a clear H$_2$-dominated atmosphere as suggested by recent observations \citep{edwards2020}. Such an observation would further help better constrain the atmospheric CH$_4$.
If future observations suggest a thin H$_2$-dominated atmosphere on LHS~1140~b, the planet is one of the best currently known targets to find potential biosignatures such as NH$_3$ or CH$_3$Cl in the atmosphere of an exoplanet in the habitable zone with a reasonable amount of JWST or ELT observation time.

\begin{acknowledgements}
This research was supported by DFG projects RA-714/7-1,
GO 2610/1-1, SCHR 1125/3-1 and RA 714/9-1. We acknowledge the support of the DFG priority programme SPP 1992 "Exploring the Diversity of Extrasolar Planets (GO 2610/2-1)". We thank Michel Dobrijevic for providing their simulated chemical profiles of Neptune and Titan and discussions on the CH$_4$ chemistry on Neptune. We also thank Billy Edwards for providing the spectral resolved HST WFC3 data of LHS~1140~b.
\end{acknowledgements}

\begin{appendix}

\section{Neptune validation} \label{app:neptune}

\begin{figure}
\centering
   \includegraphics[width=0.49\textwidth]{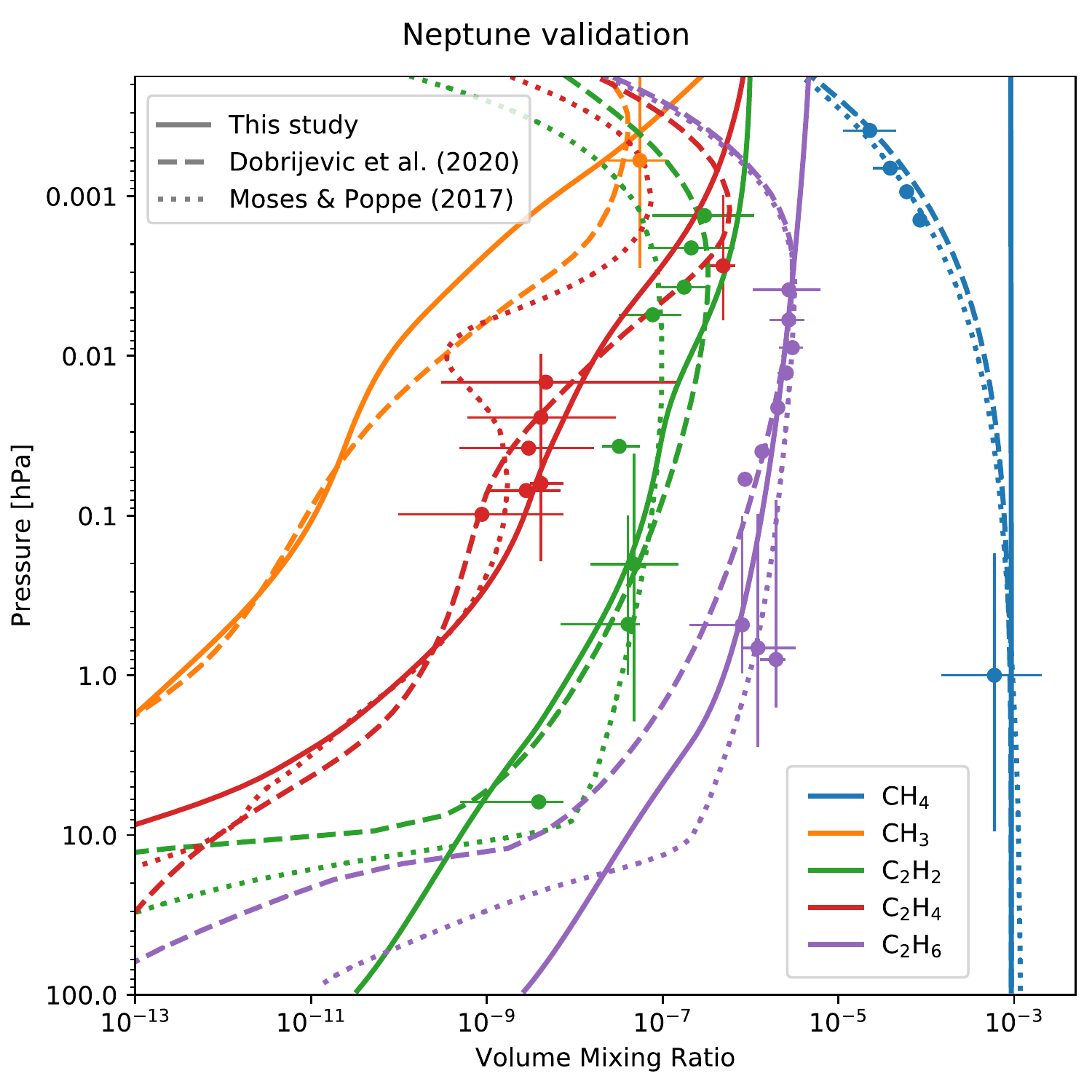}
   \caption{Neptune composition profiles for CH$_4$ (blue), CH$_3$ (orange), C$_2$H$_2$ (green), C$_2$H$_4$ (red) and C$_2$H$_6$ (purple)  calculated with our photochemistry model indicated with solid lines, compared to the model results of \citet{dobrijevic2020} with dashed lines, \citet{moses2017} (dotted lines) and a range of multiple observations (dots and corresponding error bars; see text for details).}
     \label{fig:neptune}
\end{figure}

Figure~\ref{fig:neptune} shows stratospheric composition profiles of selected species in the Neptunian atmosphere, simulated with the photochemistry model BLACKWOLF \citep{wunderlich2020}. We use the atmospheric temperature profile from \citet{fletcher2010}, inferred from infrared measurements. The profiles are compared to observations and results from \citet{dobrijevic2020}\footnote{\url{http://perso.astrophy.u-bordeaux.fr/~mdobrijevic/photochemistry}} and \citet{moses2017}. The observations are taken from numerous studies as follows: CH$_4$ from \citet{yelle1993,fletcher2010} and \citet{lellouch2015}; CH$_3$ from \citet{bezard1999}; C$_2$H$_2$, C$_2$H$_4$ and C$_2$H$_6$ from \citet{yelle1993,fletcher2010} and \citet{greathouse2011}. 

The lower boundaries at 100~hPa are set to a constant mole fraction, $f$, for He, CH$_4$, CO, CO$_2$ and H$_2$O respectively to $f_{\text{He}}$ = 0.19 \citep{williams2004}, $f_{\text{CH}_4}$ = 9.3$\times$10$^{-4}$ \citep{lellouch2015}, $f_{\text{CO}}$ = 1.1$\times$10$^{-6}$ \citep{luszcz2013}, $f_{\text{CO}_2}$ = 5 $\times$ 10$^{-10}$ \citep{feuchtgruber1999}. H$_2$ is set to be the fill gas in each layer to make up the total volume mixing ratio to unity. For all other species we assume a downward flux given by the maximum diffusion velocity, $v = K_{\text{0}}/H_0$, where $K_{\text{0}}$ is the eddy diffusion coefficient and $H_0$ the atmospheric scale height at the lower boundary. 
The eddy diffusion coefficient over height, K$_\text{{zz}}$ (in cm$^2$s$^{-1}$), are adapted from \citep{moses2018}:
\begin{equation} \label{eq:Kzz}
   K{_\text{zz}}= \begin{cases} 
        10^5 \left(\frac{0.1}{P}\right)^{0.55}, & \text{if $p \leq$ 0.5 Pa} \\
        10^4 \left(\frac{0.1}{P}\right)^{0.98}, & \text{if 0.5 Pa > $p$ > 280 Pa} \\
        400, & \text{ if $p \geq$ 280 Pa}
        \end{cases}
\end{equation}

Results suggest that the Neptunian atmosphere, as simulated by the photochemistry model, compares well both with the observations as well as with the results from \citet{dobrijevic2020} and \citet{moses2017}. Observations for pressures below 0.1~hPa however suggest a depletion of CH$_4$, which is not predicted in our model. In the stratosphere of Neptune molecular diffusion is the main process that controls the relative abundance of CH$_4$ above the methane homopause. The model version applied here includes Eddy diffusion but not molecular diffusion, consistent with an overestimation of the CH$_4$ concentrations below 0.1~hPa.

\section{Titan} \label{app:Titan}

\begin{figure}
\centering
   \includegraphics[width=0.49\textwidth]{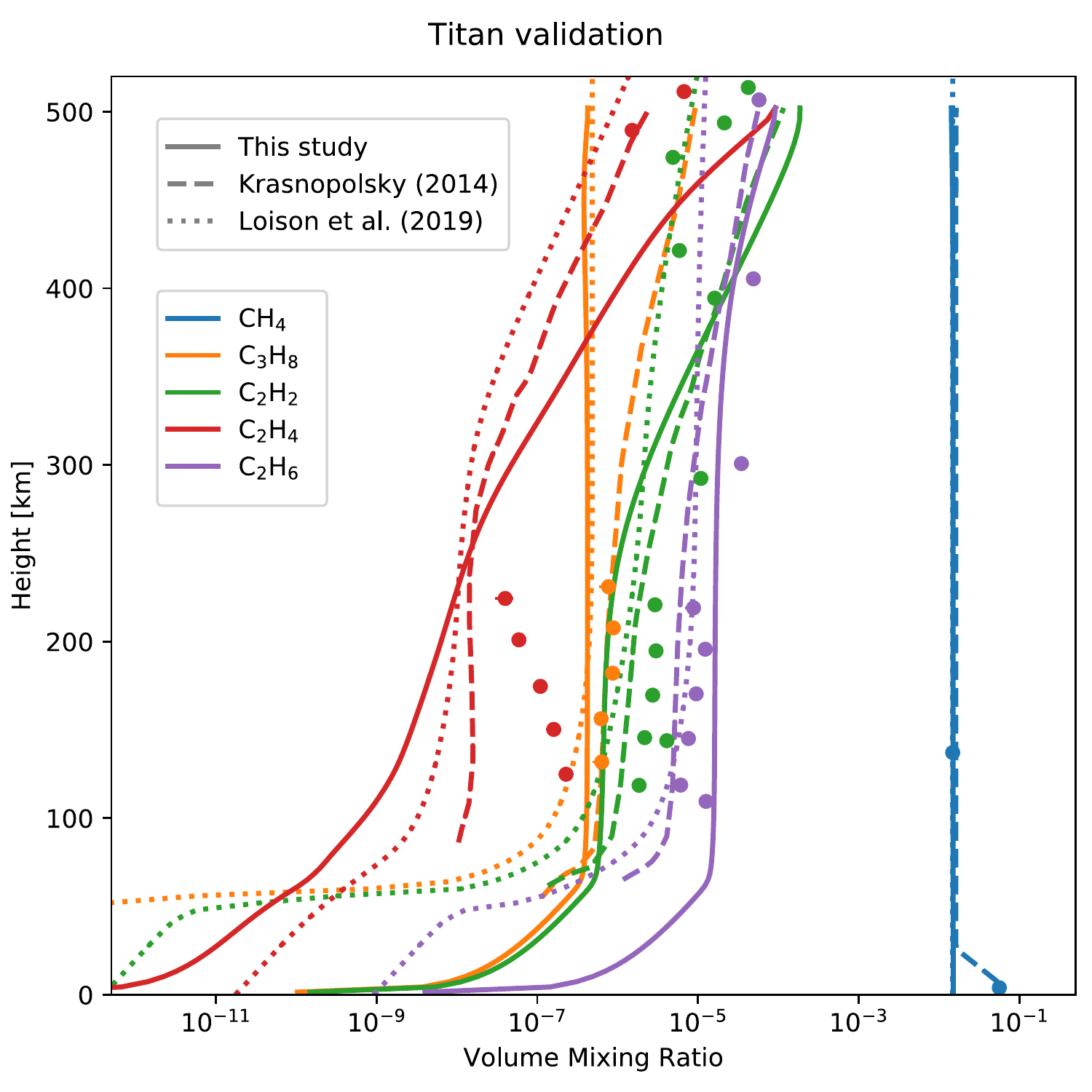}
   \caption{Titan composition profiles for CH$_4$ (blue), C$_3$H$_8$ (orange), C$_2$H$_2$ (green), C$_2$H$_4$ (red) and C$_2$H$_6$ (purple) calculated with our photochemistry model (solid lines), compared to the results from \citet{loison2019} (dashed lines), \citet{krasnopolsky2014} (dotted lines) and a range of multiple observations (dots and corresponding error bars; see text for details).}
     \label{fig:titan}
\end{figure}

Figure~\ref{fig:titan} shows the composition profiles of selected species in the atmosphere of Titan, calculated with BLACKWOLF and compared to the results from \citet{loison2019} and \citet{krasnopolsky2014}. The observations are taken from \citet{nixon2013, kutepov2013} and \citet{koskinen2011}.  

At the surface (1.45~bar) we set a constant mole fraction, $f$, for CH$_4$, CO and H$_2$ respectively to $f_{\text{CH}_4}$ = 0.015 \citep{niemann2010}, $f_{\text{CO}}$ = 4.7 $\times$ 10$^{-5}$ \citep{dekok2007} and $f_{\text{H}_2}$ = 1 $\times$ 10$^{-3}$ \citep{niemann2010}. N$_2$ is set to be the fill gas in each layer. For all other species we assume a dry deposition velocity of 0.02 cm/s. The eddy diffusion profile is taken from \citet{krasnopolsky2014}. 
The temperature profile is taken from \citet{loison2019}.

Our simulated atmosphere of Titan compares reasonably well with the results of \citet{krasnopolsky2014} and \citet{loison2019} and is consistent with the observations. Note that we simulate annual and global mean conditions with the model whereas the measurements do not represent the full range of temporal and spatial variations in the atmosphere of Titan. Profiles of latitudinal variations of the atmospheric composition of Titan are shown in \citet[e.g.][]{vinatier2010}.   %-> titan obs range: https://yly-mac.gps.caltech.edu/Titan/titan%20new/titan_Vinatier_CIRS-limb_Titan-VMRS_Icarus-2010%20copy.pdf

\section{Representation of thick hazes in GARLIC} \label{app:hazes}

\begin{figure}
\centering
   \includegraphics[width=0.49\textwidth]{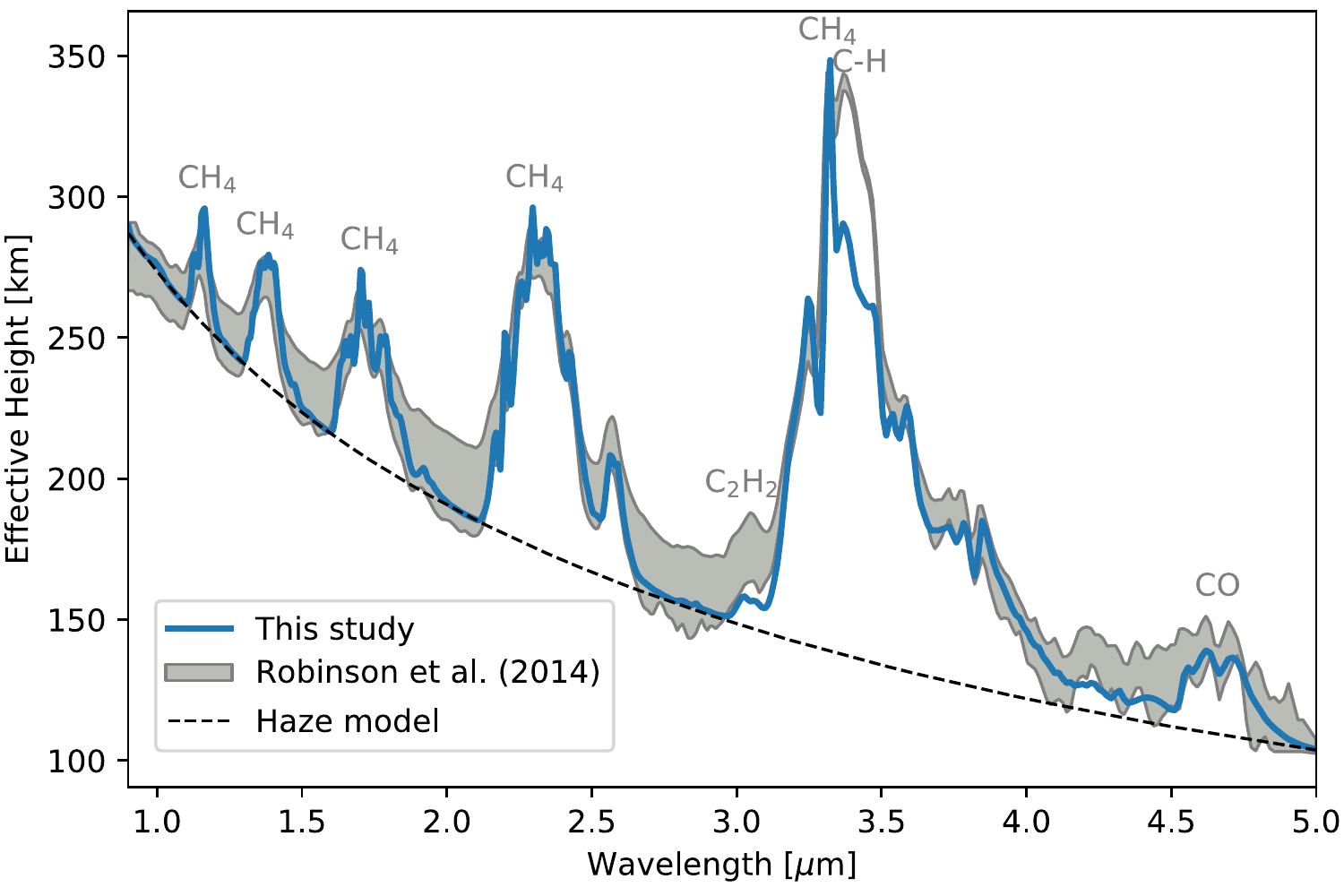}
   \caption{Transit transmission spectrum for Titan represented by effective height in km (blue solid line). The shaded region indicates deviations from four individual transit spectra taken from \citet{robinson2014}. The black dashed line shows the best fit haze model.}
     \label{fig:titan_specrum}
\end{figure}

We simulate the transmission spectrum of Titan with GARLIC using the output of BLACKWOLF (see Appendix~\ref{app:Titan}). Since the ToA for Titan in our model is at $\sim$500~km, we use the data from \citet{loison2019} to extend the atmosphere up to 1500~km. 
GARLIC represents the extinction by hazes using Eq.~(\ref{eq:sigma_aer}). We vary $\alpha$ and $\beta$ to fit the transmission spectrum from \citet{robinson2014} observed by the Visual and Infrared Mapping Spectrometer (VIMS) from \citet{brown2004} aboard the Cassini spacecraft. The best fit is presented in Fig.~\ref{fig:titan_specrum} using $\alpha = 2.6$ and $\beta = 6.0\times10^{-25}$. We use these values to simulate the impact of extinction from thick hazes in the atmosphere of LHS~1140~b. 

The CH$_4$ absorption features are reproduced well by GARLIC. We underestimate the absorption of the C-H stretching mode of aliphatic hydrocarbon chains near 3.4~$\mu$m \citep[see][]{bellucci2009,maltagliati2015}. 
This discrepancy is likely due to incomplete line lists or cross sections in the HITRAN 2016 database for several hydrocarbons such as the allyl radical \citep[C$_3$H$_5$;][]{dairene1998,desain1999}, butane \citep[C$_4$H$_{10}$;][]{abplanalp2019} and methylacetylene \citep[CH$_3$CCH;][]{abplanalp2019}. Further, our chemical network lacks some of the higher hydrocarbons for which absorption cross sections exists such as isoprene \citep[C$_5$H$_8$;][]{brauer2014}.

%some species not available cross sections
%https://aip.scitation.org/doi/pdf/10.1063/1.477425?casa_token=B0EGHsaqM8wAAAAA:ZjkIqZ8utHw5DVFM5vipo2VdkpRjL6S-6W7WkpXQxbK07-sYlXxlotk1YOfYmfHgXL5toum-icAv2Q

%or we do not include them in our model
%https://amt.copernicus.org/articles/7/3839/2014/amt-7-3839-2014.pdf

%https://pubs.rsc.org/en/content/articlepdf/2018/cp/c8cp03921f?casa_token=WhN_92zhUzwAAAAA:nKpdGuDuIcJfsllw9kq6SPQxzjfcGfftXj-Oa0wZfb-Mq0SjLyvhmU-GVNyV8qpJdusvFMWgY9X6x3IK

%https://www.sciencedirect.com/science/article/abs/pii/S0022285299978682?via%3Dihub

\end{appendix}


%% references
\bibliographystyle{aa} % style aa.bst
\bibliography{references} % your references Yourfile.bib

\end{document}